\documentclass[12pt,a4paper]{article}
\usepackage{jheppub}
\usepackage{amsmath,amsfonts,amssymb}
\usepackage{mathtools}
\usepackage{braket}
\usepackage{multirow}

\newcommand{\be}{\begin{equation}}
\newcommand{\ee}{\end{equation}}
\newcommand{\bi}{\begin{itemize}}
\newcommand{\ei}{\end{itemize}}
\newcommand{\bea}{\begin{eqnarray}}
\newcommand{\eea}{\end{eqnarray}}

\let\a=\alpha \let\b=\beta    
        \let\l=\lambda
\let\m=\mu    \let\n=\nu          
\let\s=\sigma      
    
\let\G=\Gamma    \let\L=\Lambda

\let\==\equiv

\newcommand{\Tr}{{\rm Tr}}

\newcommand{\cB}{\mathcal{B}}
\newcommand{\cC}{\mathcal{C}}

\newcommand{\cM}{\mathcal{M}}

\newcommand{\cO}{\mathcal{O}}
\newcommand{\cR}{\mathcal{R}}

\newcommand{\cV}{\mathcal{V}}

\newcommand{\half}{\tfrac{1}{2}}

\newcommand{\e}{\mathrm{e}}
\newcommand{\p}{\partial}

\newcommand{\gt}{\tilde{g}}

\newcommand{\ut}{\tilde{u}}

\title{Covariant computation of effective actions \\ in Ho\v{r}ava-Lifshitz gravity}

\author{Giulio D'Odorico,}
\author{Jan-Willem Goossens,}
\author{Frank Saueressig}
\affiliation{
Institute for Mathematics, Astrophysics and Particle Physics (IMAPP),\\
Radboud University Nijmegen, Heyendaalseweg 135, 6525 AJ Nijmegen, The Netherlands
}
\emailAdd{g.dodorico@science.ru.nl}
\emailAdd{j.goossens@science.ru.nl}
\emailAdd{f.saueressig@science.ru.nl}

\abstract{We initiate the systematic computation of the heat-kernel coefficients for Laplacian operators obeying anisotropic dispersion relations in curved spacetime. Our results correctly reproduce the limit where isotropy is restored and special anisotropic cases considered previously in the literature. Subsequently, the heat kernel is used to derive the scalar-induced one-loop effective action and beta functions of 
Ho\v{r}ava-Lifshitz gravity. We identify the Gaussian fixed point which is supposed to provide the UV completion of the theory. In the present setting, this fixed point acts as an infrared attractor for the renormalization group flow of Newton's constant and the high-energy phase of the theory is screened by a Landau pole. We comment on the consequences of these findings for the renormalizability of the theory.
	}

\begin{document}
\maketitle

%%%%%%%%%%%%%%%%%%%%%%%%%%%%%%%%%%%%%%%%%%%%%%%%%%%%%%%%%%%%%%%%%%%%%%%%%%%%%%%%%%%%%%%%%%%%

%-------------------------------------------------------------
\section{Introduction}
%-------------------------------------------------------------
Quantizing classical general relativity results in a quantum field theory that is perturbatively non-renormalizable \cite{'tHooft:1974bx,Goroff:1985sz}. This feature can readily be traced back to the negative mass-dimension of Newton's constant which entails that an infinite number of counterterms is needed 
to absorb the divergences in the loop-expansion. In terms of Wilson's modern viewpoint on renormalization this is equivalent to the fact that the Gaussian fixed point, representing the free theory, is not an ultraviolet (UV) attractor for the renormalization group (RG) flow of Newton's constant \cite{Weinberg:1980gg}. Contrary to QCD, general relativity is not asymptotically free.

Ho\v{r}ava-Lifshitz (HL) gravity \cite{Horava:2009uw} (also see \cite{Sotiriou:2009bx,Mukohyama:2010xz,Horava:2011gd,Visser:2011mf} for reviews) proposes that the UV completion of general relativity 
is provided by a different fixed point, namely an \emph{anisotropic Gaussian fixed point} (aGFP) \cite{D'Odorico:2014iha}. 
The central idea of this proposal is to treat time and space on a different footing.
At high energies, local Lorentz invariance is broken by imposing an anisotropic scaling
between space and time of the form
\be\label{scalingrel}
t \rightarrow b^{-z} \, t \, , \quad x^i \rightarrow b^{-1} \, x^i \, .
\ee
This scaling property allows the inclusion of higher-order spatial derivative terms in the action while 
retaining two time derivatives in the kinetic part. This way the theory is manifestly unitary. 
If the anisotropy parameter $z$ equals the number of spatial dimensions $d$, the mass dimension 
of Newton''s constant is zero and the coupling is power counting marginal.
This feature makes HL gravity power-counting renormalizable and constitutes an important (though not sufficient) ingredient for the theory being asymptotically free at the anisotropic Gaussian fixed point (aGFP). The
quantum properties of HL gravity beyond power-counting arguments
have recently been investigated analytically \cite{Iengo:2009ix,Rechenberger:2012dt,Benedetti:2013pya,Contillo:2013fua,Alexandre:2015fia} as well as within Monte Carlo simulations based on Causal Dynamical Triangulations \cite{Ambjorn:2010hu,Ambjorn:2013joa,Anderson:2011bj,Ambjorn:2014gsa}.

Since the anisotropic scaling \eqref{scalingrel} treats space and time in a manifestly different way, the natural choice for the fundamental degrees of freedom in HL gravity are the component fields appearing in the Arnowitt-Deser-Misner (ADM) \cite{Arnowitt:1962hi} decomposition of the spacetime metric
\footnote{We will work in the Euclidean signature. Since the foliation structure naturally defines a global time direction,
one can safely perform a Wick rotation to change the signature.}
\be
\label{ADMmet}
ds^2 = N^2 \, dt^2 + \sigma_{ij} \left( dx^i + N^i dt \right) \left(dx^j + N^j dt \right) \, . 
\ee
Here $N(t, x^i)$, $N^i(t, x^i)$ and $\sigma_{ij}(t, x^i)$ denote the lapse function,
the shift vector and the spatial metric on $d$-dimensional spatial slices of constant time $t$, respectively. Their scaling dimension is given by
\be\label{scalingdim}
\left[N\right] = 0 \, , \quad \left[\sigma_{ij}\right] = 0 \, , \quad \left[N^i\right] = z - 1 \, .  
\ee

The anisotropic scaling \eqref{scalingrel} implies that the 
symmetry group of general relativity, comprising spacetime diffeomorphisms, is reduced to
so-called foliation preserving diffeomorphisms
\be\label{fpdiff}
t \mapsto \bar{t}(t) \, , \qquad x^i \mapsto \bar{x}^i(t,x^i) \, .
\ee
The fundamental action for HL gravity then comprises all power-counting relevant and marginal interactions
that can be built from the field content and are compatible with these symmetries.  
Schematically, this action has the form
\be\label{HLaction}
S_{HL} = \frac{1}{16 \pi G} \int dt d^dx N \sqrt{\sigma} \left[ K_{ij} K^{ij} - \lambda K^2 + \cV \right] \, . 
\ee
where $K_{ij}$ 
is the extrinsic curvature (see Sec. 2) and $K$ 
denotes its trace.
The potential $\cV$ is constructed from the intrinsic curvature tensors and their spatial derivatives,
together with spatial derivatives of the component fields in (\ref{ADMmet}).

The precise form of \eqref{HLaction} depends on the version of HL gravity under consideration.
Projectable HL gravity restricts the lapse function to be independent of the spatial coordinates $x^i$
in order to match the symmetries \eqref{fpdiff}. In the non-projectable version
the lapse is allowed to depend on both the time and spatial coordinates. The ``healthy extension''
of non-projectable HL gravity \cite{Blas:2009qj,Blas:2010hb} constructs 
the potential $\cV$ from the intrinsic curvature as well as spatial derivatives of the lapse 
function. Since the stability of the theory at the quantum level
may require the terms added in the ``healthy extension'' we will also comment 
on this case in the sequel.\footnote{For a recent interesting analysis including mixed derivative terms see  \cite{Colombo:2014lta,Colombo:2015yha}.}

Despite the vast literature on HL gravity, little is known about
the quantum aspects of the theory. The reason for this lack of
understanding lies in the unfavorable interplay between
the complicated nature of the theory, including a vast set of coupling constants,
the presence of the anisotropic scaling and the complicated way the propagating 
degrees of freedom are encoded in the formulation.
One way to bypass these obstructions and address questions related to 
quantum corrections in a curved spacetime in an efficient way, is provided by
covariant heat-kernel techniques in combination with the background field method.
The key advantage of this approach is its universality: once the heat-kernel coefficients 
are known they can be used to study a multitude of physical systems. In particular
it is straightforward to compute the one-loop effective action of a theory \cite{Buchbinder:1992rb,Vassilevich:2003xt}.

The heat kernel of a second order Laplace-type differential operator $\Delta$ on
a $D$-dimensional compact manifold without boundary possesses the Seeley-deWitt (or early-time) expansion \cite{Vassilevich:2003xt,Gilkey:1995mj,Avramidi:2000bm}
\be\label{hkexp}
{\rm Tr} \, e^{-s \Delta} \simeq \frac{1}{(4\pi s)^{D/2}} \int d^Dx \sqrt{g} \, \sum_{k \ge 0} \, a_{2k} \, s^k \, ,
\ee
where the heat-kernel coefficients $a_{2k}$ are built from curvature tensors and their covariant derivatives. A similar expansion also holds for differential operators of fourth order \cite{Barvinsky:1985an,Gusynin:1988zt}.
The analogue of \eqref{hkexp} for 
differential operators
including higher-derivatives and a degenerate leading
symbol, which is the mathematical equivalent of the
differential operators naturally appearing in HL gravity, is largely uncharted territory, however.
The present work therefore serves the following purposes. Firstly, the
detailed derivation of the heat-kernel coefficients \cite{D'Odorico:2014iha} based on
off-diagonal heat-kernel techniques \cite{Anselmi:2007eq,Benedetti:2010nr} is given.
The result is extended to include terms with spatial derivatives
of the lapse function which are absent in the projectable version
of HL gravity. Subsequently, this heat kernel is used to
compute the one-loop effective action obtained from
integrating out scalar fields with anisotropic scaling relations
in an arbitrary curved background. 
The one-loop beta functions obtained in this way indeed exhibit an aGFP.
In this setting the aGFP acts as an infrared (IR) attractor while the high-energy phase
of the theory is screened by a Landau pole. In this sense the scalar-driven
gravitational renormalization group flow of HL gravity shares many features
with the one of QED. In particular, the flow is not asymptotically free.

The rest of the paper is organized as follows. 
We start by introducing our model in Sect.\ 2.
In order to facilitate
readability, the heat-kernel coefficients for differential operators
with anisotropic scaling are summarized in Sect.\ 3. Owed
to the rather technical nature, the details of their derivation
are relegated to two appendices: App.\ A contains
a general exposition of the off-diagonal heat-kernel techniques
which underlie the explicit computation of the heat-kernel coefficients
in App.\ B. The one-loop effective action together with
the beta functions for the gravitational couplings are obtained
in Sect.\ 4 and we close with a brief discussion
of our findings in Sect.\ 5.

%-------------------------------------------------------------
\section{Anisotropic gravity and matter}
\label{Sect.2}
%-------------------------------------------------------------

%-------------------------------------------------------------
\subsection{The model}
%-------------------------------------------------------------

In this work we focus on matter induced gravitational beta functions.
The matter sector of our model consists of a (massive) Lifshitz scalar (LS), with action
\be\label{scalars} 
S_{LS} =\frac{1}{2}\int dt d^d x N \sqrt{\sigma} \phi [\Delta_t + (\Delta_x)^z + m^2] \phi \,,
\ee
where
\be\label{DtDx}
%\begin{split}
\Delta_t \equiv - \frac{1}{N\sqrt{\sigma}} \, \p_t \, N^{-1} \sqrt{\sigma} \, \p_t \,,\,\,\,\,\,\,\,
\Delta_x \equiv - \frac{1}{N\sqrt{\sigma}} \, \p_i \, \sigma^{ij} N \sqrt{\sigma}  \, \partial_j \,.
%\end{split}
\ee
For isotropic scaling, $z=1$, the differential operator 
\be\label{diffop}
D^2 \equiv \Delta_t + (\Delta_x)^z
\ee
in \eqref{scalars} coincides with  the Laplace-operator $\Delta^{(D)}$ constructed from 
\eqref{ADMmet} with the shift vector set to zero
\be\label{Dlaplacian}
\left. D^2 \phi \right|_{z=1} = \left( \Delta_t + \Delta_x \right) \phi = \Delta^{(D)} \phi  \, . 
\ee 
Hence, for $z=1$ the matter sector agrees with the one of a diffeomorphism-invariant
minimally-coupled massive scalar field in a curved background spacetime.
Notice also that in the projectable case $\Delta_x$ coincides with the intrinsic Laplacian
$\Delta_x = -\sigma^{ij}\nabla_i \nabla_j$, where $\nabla_i$ 
is the covariant derivative constructed from the spatial metric $\sigma_{ij}$.

The full bare action of the theory will then be
\be\label{Sbare}
S[N,N_i, \sigma, \phi] = S_{HL}[N, N_i, \sigma] + S_{LS}[N,N_i,\sigma,\phi] \,,
\ee
where the Ho\v{r}ava-Lifshitz (HL) action is
\be 
S_{HL}= \frac{1}{16 \pi G} \int dt d^dx N \sqrt{\sigma} \left[ K_{ij} K^{ij} - \lambda K^2 + \cV \right] \,.
\ee
The extrinsic curvature is defined as 
\be\label{extcurve}
K_{ij} \equiv (2 N)^{-1} \left[\p_t \sigma_{ij} - \nabla_i N_j - \nabla_j N_i \right],
\ee
its trace being $K \equiv \sigma^{ij} K_{ij}$.
The potential term $\cV$ is built out of power-counting relevant and marginal operators.
In the projectable case, one has, in $d=2,3$
\be
\cV^{d=2} = g_0 + g_1 R + g_2 R^2, 
\ee
and
\begin{eqnarray}
\cV^{d=3} &=&  g_0 + g_1 R + g_{2,1} R^2 + g_{2,2} R_{ij}R^{ij} - g_{3,1} R \Delta_x R - g_{3,2} R_{ij} \Delta_x R^{ij} \nonumber \\
&& + g_{3,3} R^3 + g_{3,4} RR_{ij}R^{ij} + g_{3,5} R^i_jR^j_kR^k_i \, . 
\end{eqnarray}
All intrinsic curvature monomials such as $R$, $R_{ij}R^{ij}$, $\ldots$ are
built from the  induced metric $\sigma_{ij}$ on the $d$-dimensional spatial slices and indices are raised and lowered with that same metric.

In the non-projectable case the potentials include additional terms, e.g., built from spatial derivatives of the lapse function
\be
a_i \equiv \frac{1}{N} \, \p_i \, N \, .
\ee
In \cite{Blas:2009qj} for instance the following terms were added to the potential in order to give a ``healthy extension'' of HL gravity
\be
\Delta \cV = u_1 \, a_i \, a^i + u_{2,1} \, R \, \nabla_i \, a^i + u_{2,2} \, a_i \, \Delta_x \, a^i  + u_{3,1} (\Delta_x R) \,  D_i \, a^i + u_{3,2} \, a_i \, (\Delta_x)^2 \, a^i \, .
\ee
This choice limits the potential to terms which contribute to the propagator around a flat Minkowski background. In principle, the ``healthy extension'' of HL gravity should include all interaction monomials built from the component fields in \eqref{ADMmet} which are compatible with the symmetries \eqref{fpdiff}. In total this setup then includes $\cO(100)$ coupling 
constants, leading to a highly involved theory.

%-------------------------------------------------------------
\subsection{Effective actions from the heat kernel}
%-------------------------------------------------------------
Starting from an action functional $S[\phi]$, the one-loop contribution to the effective action (EA) takes the form \cite{Buchbinder:1992rb,Avramidi:2000bm}\footnote{Technically one should divide the Hessian $\frac{\delta^2 S}{\delta\phi \delta\phi}$ by a suitable mass scale $\mu$ 
so that the argument of the logarithm is dimensionless. At this stage, this scale will be left implicit. It will appear as the renormalization group scale later on.}
\begin{equation}
\label{gamma1}
\Gamma_1 = \frac{1}{2} \mathrm{Tr} \, \mathrm{log} \, \left[ \frac{\delta^2 S}{\delta\phi \delta\phi} \right] \,.
\end{equation}
The one-loop effective action is then given by $\Gamma=S+\Gamma_1$.
Typically, the functional trace (\ref{gamma1}) contains divergences
which are absorbed into the counterterms contained in $S$. These
divergences are crucial for the renormalization of the theory since they encode the beta functions governing the energy dependence of the coupling constants. A very convenient way to isolate these divergences in curved spacetime
is by means of the heat kernel. This approach preserves the covariance
of the effective action with respect to the underlying symmetry group. We will illustrate this technique in the remainder of this subsection before applying it to HL gravity in Sect.\ \ref{Sect.4}.

The functional trace of $\mathrm{log}\, \cal O$, with $\cal O$ an operator 
(e.g. the Laplacian operator on a $D$-dimensional Riemannian manifold)
can be expressed through the eigenvalues $\lambda_n$ of $\cal O$ as
\begin{equation}\label{zetafunct}
\frac{1}{2} \mathrm{Tr} \, \mathrm{log} \, {\cal O} = \frac{1}{2}\sum_n \mathrm{log}\,\lambda_n = -\frac{1}{2}\sum_n \left. \frac{d}{ds} \lambda_n^{-s} \right|_{s=0}
\equiv -\frac{1}{2} \left. \frac{d}{ds} \, \zeta_{\cal O} (s) \right|_{s=0} \,.
\end{equation}
The last equality defines the zeta-function of $\cal O$. The zeta-function can be analytically continued to the whole complex plane which gives a way to define a finite EA. This is the idea of zeta-function regularization \cite{Hawking:1976ja}.
There is however another way of expressing the functional trace.
The sum over eigenvalues
appearing in \eqref{zetafunct} can formally be written as
\begin{equation}
\sum_n \lambda_n^{-s} = \frac{1}{\Gamma(s)}\int_0 ^{\infty} dt \, t^{s-1} \sum_n e^{-t \lambda_n} =
\frac{1}{\Gamma(s)}\int_0 ^{\infty} dt \, t^{s-1} {\rm Tr} e^{-t \cal O} \, . 
\end{equation}
In this way one obtains the formal equivalence
\begin{equation}
\label{1loop}
\frac{1}{2} {\rm Tr}\, \mathrm{log}\, {\cal O} = - \frac{1}{2} \int_0 ^{\infty} \frac{dt}{t} {\rm Tr} e^{-t \cal O}
\end{equation}
which is exact up to a field-independent infinite constant which does not contribute to the EA. The relation \eqref{1loop} reduces the computation of the one-loop effective action to determining the quantity
\be\label{defheat} 
K_{\cal O}(s) \equiv {\rm Tr} \, e^{-s \cal O} \, .
\ee
Formally, $K_{\cal O}(s)$ satisfies the heat equation
\be 
\partial_s K_{\cal O}(s) + {\cal O} K_{\cal O}(s) =0 \,,
\ee
with boundary condition $K_{\cal O}(0)=1$.
For this reason $K_{\cal O}(s)$ is called the heat kernel of the operator ${\cal O}$.

There exist powerful methods to compute the heat kernel in terms of invariants on the manifold \cite{Vassilevich:2003xt,Gilkey:1995mj,Avramidi:2000bm}.
The most important one for computing the effective action is the Seeley-deWitt expansion which constitutes an asymptotic expansion of \eqref{defheat}
for $s\to 0$. For the second order Laplace operator $\Delta^{(D)}$
this expansion reads
\begin{eqnarray}
\label{heatkernel}
{\rm Tr} \, e^{-s\Delta^{(D)}} 
&=& \left(4\pi \right)^{-\frac{D}{2}} s^{-\frac{D}{2}}
\int d^Dx \sqrt{g} \sum_{n \geq 0,i}  s^{n} \, a_{2n,i} \, {\cal R}_{2n}^{(i)}
\end{eqnarray}
where the $a_{2n,i}$ are 
$D$-independent numerical coefficients and the ${\cal R}_{2n}^{(i)}$ 
span a basis of curvature monomials built from $2n$ covariant derivatives.
The diffusion time $s$ has mass dimension $-2$, so each term within the sum is dimensionless and the rhs has mass dimension zero.

When we plug the expansion \eqref{heatkernel} into (\ref{1loop}), we see that there are UV divergences associated with the lower boundary of the 
integration interval. These divergences arise from monomials which come with a negative or zero power of $s$. For monomials multiplied by positive powers of $s$ the integral vanishes at the lower boundary indicating that these terms do not give rise to UV divergences. The former class of interaction monomials precisely correspond to the relevant ($s < 0$) and marginal ($s = 0$) operators of the theory which require counterterms provided by the bare action.

For HL gravity, the operator we are interested in is the anisotropic Laplacian \eqref{diffop} which constitutes the Hessian for the action of the Lifshitz scalar \eqref{scalars}. 
Based on the Seeley-deWitt expansion for this anisotropic operator, we can immediately construct the matter-induced effective action at the one-loop level and obtain the matter-induced beta functions for the relevant and marginal coupling constants of the theory.

%-------------------------------------------------------------
\section{Heat-kernel results for anisotropic differential operators}
\label{sect.3}
%-------------------------------------------------------------
Based on dimensional analysis, the Seeley-deWitt expansion for the
anisotropic differential operator \eqref{diffop} must take the form
\begin{equation}
%\begin{split}
{\rm Tr} \, e^{-s D^2} = (4\pi)^{-\frac{d+1}{2}} \, s^{-\frac{1+d/z}{2}} \, \int dt d^dx \, N \sqrt{\sigma} \sum_{l,m,n\ge 0} 
s^{\frac{l(1-z)}{2z} + \frac{m}{2} + \frac{n}{2z}} \, {\rm tr} \, {\bf b}_{l,m,n} \, .
%\end{split}
\label{mastermaster}
\end{equation}
Here the number of shift vectors $N^i$, time derivatives, and spatial derivatives contained in the coefficient ${\bf b}_{l,m,n}$ is given by
$l$, $m$, and $n$, respectively and tr denotes a trace over internal indices.
Comparing eqs.\ \eqref{heatkernel} and \eqref{mastermaster} one first notices a change in the overall $s$-dependence of the rhs as well as the appearance of the anisotropy parameter $z$ in the factors $s$ multiplying interaction monomials including the lapse $N^i$ and spatial derivatives. This change of scaling is understood as follows. With our scaling conventions,
time derivatives get mass dimension $z$, while spatial ones keep their standard dimension 1, so that $s$ has dimension $-2z$.
Then, as before, the overall scaling has to compensate the dimension of the integration measure,
while the powers of $s$ within the sum compensate those of the operators,
so that all terms appearing on the rhs are dimensionless.
\begin{table}[t]
	\renewcommand{\arraystretch}{1.6}
	\begin{center}
		\begin{tabular}{|c||c|c|c|c|c|c|c|c|}
			\hline
			& & \multicolumn{3}{|c|}{$d=2$} & \multicolumn{4}{|c|}{$d=3$} \\ \hline 	 
			& dim	& $a_{2n}$ & $z=2$ & $z = 3$ & $a_{2n}$ & $z=2$ & $z = 3$ & $z=4$  \\ \hline \hline
			$K^2$ & $\frac{d-z}{2z} $ & $\frac{1}{6}$ & $\frac{\sqrt{\pi}}{16}$ & $\frac{1}{36}  \Gamma\left(\frac{1}{3}\right)$ & $\frac{1}{6}$ & $\frac{2\Gamma\left( \frac{3}{4} \right)}{15\sqrt{\pi}}$ & $\frac{1}{15}$ & $\frac{\Gamma\left( \frac{3}{8} \right)}{30\sqrt{\pi}}$ \\
			$K_{ij}K^{ij}$ & $\frac{d-z}{2z}$ & $- \frac{1}{6}$ & $- \frac{\sqrt{\pi}}{8}$ & $-\frac{1}{9}  \Gamma\left(\frac{1}{3}\right)$ & $-\frac{1}{6}$  & $-\frac{7\Gamma\left( \frac{3}{4} \right)}{30\sqrt{\pi}}$ & $-\frac{1}{5}$ & $- \frac{11\Gamma\left( \frac{3}{8} \right)}{60\sqrt{\pi}}$ \\ \hline
			$1$  & $\frac{d+z}{2z}$ & $1$ & $\frac{\sqrt{\pi}}{2}$ & $\Gamma\left(\frac{4}{3}\right)$ & $1$ & $\frac{4 \Gamma\left(\frac{7}{4}\right)}{3\sqrt{\pi}}   $ & $\frac{2}{3}$ & $\frac{4 \Gamma\left(\frac{11}{8}\right)}{3\sqrt{\pi}} $ \\
			$R$  &  $\frac{d-2+z}{2z}$ & $\frac{1}{6}$ & $\frac{1}{6}$ & $\frac{1}{6}$ & $\frac{1}{6}$ & $\frac{ \Gamma\left(\frac{5}{4}\right)}{3\sqrt{\pi}} $ & $\frac{ \Gamma\left(\frac{7}{6}\right)}{3\sqrt{\pi}} $ & $\frac{ \Gamma\left(\frac{9}{8}\right)}{3\sqrt{\pi}} $  \\
			$R^2$ & $\frac{d-4+z}{2z}$ & $\frac{1}{60}$ & $0$ & $0$ & $\frac{1}{120}$ & $\frac{ \Gamma\left(\frac{5}{4}\right)}{120\sqrt{\pi}} $ & $\frac{ \Gamma\left(\frac{7}{6}\right)}{120\sqrt{\pi}} $ & $\frac{ \Gamma\left(\frac{9}{8}\right)}{120\sqrt{\pi}} $ \\
			$R_{ij}R^{ij}$ & $\frac{d-4+z}{2z}$ & $-$ & $-$ & $-$ & $\frac{1}{60}$ & $\frac{ \Gamma\left(\frac{5}{4}\right)}{60\sqrt{\pi}} $ & $\frac{ \Gamma\left(\frac{7}{6}\right)}{60\sqrt{\pi}} $ & $\frac{ \Gamma\left(\frac{9}{8}\right)}{60\sqrt{\pi}} $  \\
			$-R\Delta_xR$ & $\frac{d-6+z}{2z}$ & $-$ & $-$ & $-$ & $\frac{1}{336}$ &  $-\frac{ \Gamma\left(\frac{5}{4}\right)}{168\sqrt{\pi}} $ & $-\frac{1}{672}$ & $-\frac{ \Gamma\left(\frac{5}{8}\right)}{672\sqrt{\pi}} $ \\
			$-R_{ij}\Delta_x R^{ij}$ & $\frac{d-6+z}{2z}$ & $-$ & $-$ & $-$ & $\frac{1}{840}$ &  $-\frac{ \Gamma\left(\frac{5}{4}\right)}{420\sqrt{\pi}} $ & $-\frac{1}{1680}$ & $-\frac{ \Gamma\left(\frac{5}{8}\right)}{1680\sqrt{\pi}} $ \\
			$R^3$ & $\frac{d-6+z}{2z}$ & $\frac{1}{756}$ & $-2$ & $0$ & $- \frac{1}{560}$ & $\frac{ \Gamma\left(\frac{5}{4}\right)}{280\sqrt{\pi}} $ & $\frac{1}{1120}$ & $\frac{ \Gamma\left(\frac{5}{8}\right)}{1120\sqrt{\pi}} $ \\
			$R R_{ij} R^{ij}$ & $\frac{d-6+z}{2z}$ & $-$ & $-$ & $-$ & $\frac{1}{105}$ & $-\frac{2 \Gamma\left(\frac{5}{4}\right)}{105\sqrt{\pi}} $ & $-\frac{1}{210}$ & $-\frac{ \Gamma\left(\frac{5}{8}\right)}{210\sqrt{\pi}} $ \\
			$R^i_j R^j_k R^k_i$ & $\frac{d-6+z}{2z}$ & $-$ & $-$ & $-$ & $- \frac{1}{180}$ &  $\frac{ \Gamma\left(\frac{5}{4}\right)}{90\sqrt{\pi}} $ & $\frac{1}{360}$ & $\frac{ \Gamma\left(\frac{5}{8}\right)}{360\sqrt{\pi}} $ \\ \hline
			$a^2$ & $\frac{d-2+z }{2z}$ & $0$ & $-\frac{13}{12}$ &  $-\frac{13}{6}$ & $0$ & $-\frac{13\Gamma\left(\frac{5}{4}\right)}{9\sqrt{\pi}} \,  $ & $-\frac{26\Gamma\left(\frac{7}{6}\right)}{9\sqrt{\pi}}  $ & $-\frac{13 \Gamma\left(\frac{9}{8}\right)}{3\sqrt{\pi}}   $ \\ \hline \hline
		\end{tabular}
		\caption{\label{table} Heat-kernel coefficients for the anisotropic differential operator $D^2$ multiplying the curvature invariants given in the first column. The second column indicates that the corresponding term scales as $s^{-{\rm dim}}$. The heat-kernel coefficients for interaction terms typically considered in the framework of HL gravity in $d=2$ and $d=3$ are given in columns 3 to 5 and 6 to 10, respectively. The dash implies that the corresponding monomial does not appear in the basis of curvature monomials in the given dimension. The coefficients for $z=1$ agree with the heat-kernel coefficients of the isotropic Laplace operator of second order on a manifold with dimension $d=2$ and $d=3$.}
	\end{center}
\end{table}

In App.\ \ref{App.B} we explicitly compute the  coefficients ${\bf b}_{l,m,n}$ 
for all interaction monomials containing two time derivatives, two spatial derivatives as well as for all intrinsic curvature monomials $\cR^{(i)}_{2n}$ constructed from the metric on the spatial slices $\sigma_{ij}$. Restricted to this (rather general) subclass of terms, the general formula \eqref{mastermaster} becomes
\begin{equation}
\begin{split}
{\rm Tr} \, e^{-s D^2} \simeq & \,  \left(4\pi \right)^{-\frac{d+1}{2}}  s^{-\frac{1+d/z}{2}} \, \int dt d^dx N \sqrt{\sigma} \, \times \\
& \, \bigg[ \frac{s}{6} \left( e_1 \, K^2 + e_2 \, K_{ij} K^{ij} \right) 
+  \sum_{n \ge 0 } \, s^{\frac{n}{z}} \,  b_n \, 
\sum_i a_{2n,i} \, {\cal R}_{2n}^{(i)} \, 
+ s^{\frac{1}{z}} \, c_1 \, a_i a^i \, \bigg]   \, .
\end{split}
\label{master}
\end{equation}
Here the $a_{2n,i}$ are the standard Seeley-deWitt coefficients associated with the intrinsic curvature monomials ${\cal R}_{2n}^{(i)}$ in the expansion \eqref{heatkernel} (see \cite{Groh:2011dw} for an explicit list). The coefficients $e_i$ and $b_n$ depend on $d$ and $z$ and encode the modifications induced by the anisotropic scaling of the differential operator. We now discuss the three sectors appearing in \eqref{master}. The coefficients most relevant for HL gravity are collected in Tab.\ \ref{table}.\footnote{The coefficients $a_{2n}$ stated in the third and sixth column of the table explicitly depend on the dimension $d$. This is owed to the fact that the results are given for ``manifolds of low dimension'', where the general basis of geometrical invariants is degenerate. Eliminating the redundant basis elements by imposing the vanishing of the Weyl tensor ($d=3$) and the vanishing of the Weyl tensor together with the Einstein condition ($d=2$) generates this $d$-dependence which is absent in the general heat-kernel expansion.} \\

\noindent
{\bf Extrinsic curvatures} \\
The coefficients $e_1$ and $e_2$ associated with the kinetic terms built from the extrinsic curvature \eqref{extcurve} are computed in App.\ \ref{sect.B3} and read
\be\label{ee}
e_1(d,z) = \frac{d-z+3}{d+2} \frac{\Gamma(\tfrac{d}{2z})}{z \Gamma(\tfrac{d}{2})} \, , \qquad
e_2(d,z) = - \frac{d+2z}{d+2} \frac{\Gamma(\tfrac{d}{2z})}{z \Gamma(\tfrac{d}{2})} \, . 
\ee
For $z=1$ these equations reduce to $e_1|_{z=1} = 1$ and $e_2|_{z=1} = -1$. This agrees with the heat-kernel coefficient $a_2$ of $\Delta^{(D)}$ expressed in terms of intrinsic and extrinsic curvatures and provides a non-trivial check of our result. At criticality, $d=z$, the coefficients \eqref{ee} satisfy the relation
\be
e_1|_{z=d} = - \frac{1}{d} \, e_2|_{z=d} \, . 
\ee
As we will see in the next section, it is this relation which underlies the fixed point structure of HL gravity.\\

\noindent
{\bf Intrinsic curvatures} \\
The coefficients $b_n$ are defined through the integral representation \eqref{eqb46} derived in App.\ \ref{App:B3}. This integral can be evaluated via Mellin transform techniques giving
\begin{eqnarray}\label{eq34}
& b_n(d,z)  =  \frac{ \Gamma(\tfrac{d-2n}{2z}+1)}{\Gamma(\tfrac{d-2n}{2}+1)} \, , \quad & 0 \le n \le \lfloor d/2 \rfloor \, , \\ \label{bngreaterdover2}
& b_n(d,z)  =  \frac{(-1)^{k}}{\Gamma(d/2-n+k)} \, \int_0^\infty dx \, x^{d/2-n+k-1} \, \left( \p_x \right)^k \e^{-x^z}  \, , \quad & n > \lfloor d/2 \rfloor \, . 
\end{eqnarray}
Here $k = n+1 - \lfloor d/2 \rfloor$ and the symbol $\lfloor x \rfloor$ denotes the floor function, namely the largest integer not greater than $x$.
For $d$ even the integral in eq.\ \eqref{bngreaterdover2} can be evaluated explicitly, yielding
\be
b_n(d,z) = (-1)^{n-d/2+n/z-d/(2z)} \, \frac{\Gamma\left(\frac{2n-d}{2}+1\right)}{\Gamma\left(\frac{2n -d}{2z} +1\right)} \, \delta_{n/z - d/(2z), q} \, , \qquad q \in \mathbb{N} \, . 
\ee
Note that the Kronecker-delta puts severe constraints on the intrinsic curvature terms that appear in the heat-kernel expansion.

At this stage the following comment on the convergence of the integral \eqref{bngreaterdover2} is in order. Due to the exponential suppression of the integrand at the upper boundary of the integration interval, possible divergences can arise at the lower boundary only. For $z$ being integer, the integrand is given by a finite polynomial times the exponential. The leading contribution for $x \to 0$ is proportional to  $x^{d/2-\lfloor d/2\rfloor}$ which is manifestly non-negative. Thus for $z \in \{\mathbb{N}/0\}$ the integral converges and the $b_n(d,z)$ are finite. When $z > 0$ is non-integer, however, the leading contribution for $x \to 0$ can acquire an exponent which is smaller or equal to $-1$. Acting with $k$ derivatives on the exponential, the term which dominates for $x\to 0$ is of the form
\be
\frac{d^k}{dx^k} e^{-x^z} = -\left( \prod_{l=0}^{k-1} (z-l)\right) x^{z-k} \e^{-x^z} + \ldots \, , 
\ee  
where the dots stand for all terms with a higher power of $x$.
Thus \eqref{bngreaterdover2} becomes
\be
b_n(d,z) = \frac{(-1)^{k+1}}{\Gamma(d/2-n+k)} \left( \prod_{l=0}^{k} (z-l)\right)\, \int_0^\infty dx \, \left( x^{d/2-\lfloor d/2 \rfloor+z-k} \e^{-x^z}  \, + \ldots \right) , 
\ee
where again the dots denote terms which are subleading at the lower integration boundary. The integral is convergent   
as long as $d/2-\lfloor d/2 \rfloor -k+1+z > 0$, or, using the definition of $k$, $d/2 -n+z>0$. 
This establishes that for non-integer $z$ the heat-kernel coefficients
$b_n(d,z)$ for $n < d/2+z$ are finite. For $n \ge d/2 + z$, the $b_n(d,z)$
diverge, however. This demonstrates that the operator $D^2$ with non-integer values $z$ does not possess 
a well-defined Seeley-deWitt expansion due to the divergence of the coefficients $b_n(d,z)$ with  $n \ge d/2 + z$.\\

\noindent
{\bf Terms containing spatial derivatives of the lapse function} \\ 
The coefficient $c_1$ is computed in App.\ \ref{App:B4} and reads
\be\label{e3}
c_1(d,z) = - \frac{13}{6} \, \frac{z-1}{d} \, \frac{\Gamma\left(\frac{d-2}{2z}+1\right)}{\Gamma\left(\frac{d}{2}\right)} \, . 
\ee
Notably, $c_1$ is proportional to $z-1$ and thus vanishes in the isotropic case. This is consistent with the fact that for $z=1$ the operator $D^2$ coincides with the standard Laplace operator whose heat-kernel expansion can not contain terms which break the invariance under diffeomorphisms. At criticality $c_1$ reduces to
\be
c_1|_{z=d} = - \frac{13}{6} \, \frac{d-1}{d} \, \frac{\Gamma\left(\frac{3}{2}- \frac{1}{d}\right)}{\Gamma\left(\frac{d}{2}\right)} \, .
\ee

The computation of $c_1$ serves as an illustration that the off-diagonal heat-kernel methods developed in this work are also capable of determining heat-kernel coefficients relevant for the ``healthy extension'' of HL gravity. Comparing the complexity of the derivations in App.\ \ref{sect.B3} and \ref{App:B4} reveals, however, that
finding the heat-kernel coefficients for non-projectable HL gravity is computationally much more demanding than in the projectable case.
Determining all heat-kernel coefficients relevant for the ``healthy extension'' will require a significant computational effort which is beyond the scope of the present paper.

We close this section with the following remark. We explicitly verified
that the isotropic limit $z=1$ of our heat-kernel coefficients correctly reproduces the expansion \eqref{heatkernel} written in terms of ADM fields.  Moreover, our formulas agree with special cases for $z=2$ and $d=1$ studied in \cite{Ignat:2013aeh}, the results for the conformal anomaly in $z=2, d=2$ obtained in  \cite{Baggio:2011ha} and the scaling analysis for anisotropic Laplace operators in flat space \cite{Mamiya:2013wqa}. This match provides a highly non-trivial test for the results obtained in this work.

%-------------------------------------------------------------
\section{One-loop effective action and beta functions}
\label{Sect.4}
%-------------------------------------------------------------
We now use the Seeley-deWitt expansion \eqref{master} to construct
the one-loop effective action and beta functions for the gravitational
sector of HL gravity resulting from integrating out the
anisotropic scalar field \eqref{scalars}.
%-------------------------------------------------------------
\subsection{General beta functions}
%-------------------------------------------------------------
Applying the general formulas \eqref{gamma1} and \eqref{1loop} to
the scalar action \eqref{scalars}, the resulting one-loop contribution
to the effective action is
\be
\Gamma_1 = -\frac{1}{2}\int_0^\infty \frac{ds}{s} \, e^{-s m^2} \, \Tr \left[ e^{-sD^2} \right] \, . 
\ee
Evaluating the operator trace via the Seeley-deWitt expansion \eqref{master}, one obtains
 \be
 \begin{split}
 \Gamma_1 = & \, -\frac{\left(4\pi \right)^{-\frac{d+1}{2}}}{2} \int dt d^dx N \sqrt{\sigma}\, \int_0^\infty \frac{ds}{s} \, e^{-s m^2} \, s^{-\frac{1+d/z}{2}} \\ & 
 \times  \, 
 \bigg[ \tfrac{s}{6} \left( e_1 \, K^2 + e_2 \, K_{ij} K^{ij} \right) + \sum_{n \ge 0 } \, s^{n/z} \,  b_n \, 
 \sum_i a_{2n,i} {\cal R}_{2n}^{(i)} + s^{\frac{1}{z}} \, c_1 \, a_i a^i + \ldots \bigg]
 \end{split}
 \ee
where the dots indicate further terms containing spatial derivatives of the lapse function whose heat-kernel coefficients are unknown. UV divergences
are linked to the lower boundary of the $s$-integration. In order to make them visible, we introduce a UV cutoff $\L$,
so the lower limit is now $\L^{-2}$, and integrate from the cutoff scale $\L$ down to some renormalization
scale $\mu$. 

It is now straightforward to determine the divergent parts of $\Gamma_1$. In the massless case, $m=0$ these are given by
\be
\begin{split}
\Gamma_{1,\L} =& -\frac{(4\pi)^{-\frac{d+1}{2}}}{2}\int dtd^dx N \sqrt{\s} \bigg[
\frac{(\L^{\frac{d}{z}-1} - \mu^{\frac{d}{z}-1} )}{3(\frac{d}{z}-1)}(e_1 K^2 + e_2 K_{ij}K^{ij})   \\
&  +\sum_{n = 0}^{\lfloor (d+z)/2 \rfloor} \frac{2z}{d-2n+z} \left(\L^{\frac{d-2n+z}{z}} - \mu^{\frac{d-2n+z}{z}} \right) \, b_n \,  
\sum_i a_{2n,i} \, {\cal R}_{2n}^{(i)}  \\
&  + \frac{2z}{d-2+z} \left(\L^{\frac{d-2+z}{z}} - \mu^{\frac{d-2+z}{z}} \right) \, c_1 \, a_i a^i + \ldots \bigg]
\label{WLM}
\end{split}
\ee
subject to the substitutions
\be\label{eq44}
\begin{split}
\left( \tfrac{d}{z} - 1 \right)^{-1} \left(\L^{\frac{d}{z}-1} - \mu^{\frac{d}{z}-1} \right) 	& \mapsto \log\left(\Lambda/\mu\right) \, , \qquad \mbox{if} \quad d=z \, , \\
	  z \left(d-2n+z \right)^{-1} \left(\L^{\frac{d-2n+z}{z}} - \mu^{\frac{d-2n+z}{z}} \right) & \mapsto \log\left(\Lambda/\mu\right) \, , \qquad \mbox{if} \quad d + z = 2n \, 
\end{split}
\ee 
applicable when a term in the heat-kernel expansion is independent of $s$ or, equivalently, if the interaction monomial is power-counting marginal. In the massive case, $m \not = 0$, the $s$-integrals give rise to incomplete Gamma functions, $\Gamma(a,b) \equiv \int_b^\infty dt \, t^{a-1} \, e^{-t}$. The analogue of \eqref{WLM} then reads
\begin{eqnarray}
\Gamma_{1,\L} &=& -\frac{(4\pi)^{-\frac{d+1}{2}}}{2}\int dtd^dx N \sqrt{\s} \bigg[
\tfrac{1}{6} \left(\G(\tfrac{1-d/z}{2},\tfrac{m^2}{\L^{2}}) - \G(\tfrac{1-d/z}{2},\tfrac{m^2}{\m^{2}})\right)(e_1 K^2 + e_2 K_{ij}K^{ij})   \nonumber \\
&&   +\sum_{n = 0}^{\lfloor (d+z)/2 \rfloor} z \left( \G(\tfrac{2n-d-z}{2z},\tfrac{m^2}{\L^{2}}) - \G(\tfrac{2n-d-z}{2z},\tfrac{m^2}{\m^{2}})\right) b_n \sum_i a_{2n,i} \, {\cal R}_{2n}^{(i)}  \nonumber \\ \label{EffMass}
&& + z \left( \G(\tfrac{2-d-z}{2z},\tfrac{m^2}{\L^{2}}) - \G(\tfrac{2-d-z}{2z},\tfrac{m^2}{\m^{2}})\right) \, c_1 \, a_i a^i + \ldots \bigg] \, . 
\end{eqnarray}

In the next step the UV divergences are reabsorbed by the counterterms present in the bare action \eqref{Sbare}. The (scale-dependent) renormalized couplings for the massive case can then be read off from \eqref{EffMass}. For the couplings appearing in projectable HL gravity
\be\label{gren}
\begin{split}
	\frac{1}{16\pi G_R(\mu)} = & \frac{1}{16\pi G} +  \frac{1}{(4\pi)^{(d+1)/2}} \,  \frac{e_2}{12} \,\G(\tfrac{1-d/z}{2},\tfrac{m^2}{\m^{2}}) \, , \\
	-\frac{\l_R(\mu)}{16\pi G_R(\mu)} = &  -\frac{\l}{16\pi G} +\frac{1}{(4\pi)^{(d+1)/2}} \, \frac{e_1}{12} \,  \G(\tfrac{1-d/z}{2},\tfrac{m^2}{\m^{2}}) \, , \\
	\frac{g_{n,i,R}(\mu)}{16\pi G_R(\mu)} = & \frac{g_{n,i}}{16\pi G} +\frac{1}{(4\pi)^{(d+1)/2}} \, z \, b_n \, a_{2n,i} \,  \G(\tfrac{2n-d-z}{2z},\tfrac{m^2}{\m^{2}}) \, .  \\
\end{split}
\ee
Here $G, \lambda$, and $g_{n,i}$ are the values of the renormalized couplings at $\mu = 0$, which constitute free parameters of the theory. 
The renormalized coupling constants in the ``healthy extension'' follow the same pattern as the $g_{n,i,R}(\mu)$. In particular,
\be\label{uren}
\frac{u_{1,R}(\mu)}{16\pi G_R(\mu)} = \frac{u_1}{16\pi G} +
\frac{1}{(4\pi)^{(d+1)/2}} \, 
z \, c_1 \, \G(\tfrac{2-d-z}{2z},\tfrac{m^2}{\m^{2}}) \, . 
\ee
The scale-dependent renormalized coupling constants of the massless
case are found in an analogous way either by reading them off directly from 
\eqref{WLM} or taking the limit $m \rightarrow 0$ in eqs.\ \eqref{gren} and \eqref{uren}. In this limit the incomplete Gamma-functions simplify to a power-law running in $\mu$ if the corresponding coupling is relevant and a logarithmic $\mu$-dependence for marginal couplings. Since the subsequent discussion involves only renormalized couplings, we will from now on drop the subscript $R$ for ease of notation.

The scale-dependence of the renormalized coupling constants is conveniently encoded in the beta functions of a theory. For this purpose, we use the renormalization group scale $\mu$ to construct the
 dimensionless couplings 
\be
\begin{split}
g(\mu) \equiv G(\mu) \, \mu^{2\eta} \, , \qquad 
\gt_{n,i}(\mu) \equiv g_{n,i}(\mu) \, \mu^{-2\xi_n} \, , \qquad
\ut_{n,i}(\mu) \equiv u_{n,i}(\mu) \, \mu^{-2\xi_n} \, ,
\end{split}
\ee
where the mass-dimensions are determined through
\be\label{scalingdims}
\eta\equiv\frac{d}{2z} - \frac{1}{2} \, , \qquad \xi_n \equiv 1-\frac{n}{z} \, .
\ee
The beta functions $\mu\p_\mu \tilde{g} \equiv \beta_{\tilde{g}}$
are obtained by re-expressing eqs.\ \eqref{gren} and \eqref{uren} in terms of the dimensionless couplings and taking the logarithmic $\mu$-derivative. Explicitly, 
\be\label{betamass}
\begin{split}
	\b_g = & \,  2 \, \eta \, g-\frac{2}{3} \, (4\pi)^{-(d-1)/2} \, e^{-\m^{-2}m^2} \, e_2 \, g^2, \\
	\b_\l =  & \, -\frac{2}{3} \, (4\pi)^{-(d-1)/2} \, e^{-\m^{-2}m^2} \left(e_1 + \l e_2 \right) \, g \, ,\\
	\b_{\gt_{n,i}} =  & \, -2 \, \xi_n \, \gt_{n,i} + g \, (4\pi)^{-(d-1)/2} \,   e^{-\m^{-2}m^2} \left(4 \, b_n \, a_{2n,i} - \tfrac{2}{3} \, \gt_{n,i} \, e_2 \right) \, , \\
	\b_{\ut_{1}} =  & \, -2 \, \xi_1 \, \ut_{1} + g \, (4\pi)^{-(d-1)/2} \,   e^{-\m^{-2}m^2} \left(4 \, c_1 - \tfrac{2}{3} \, \ut_{1} \, e_2 \right) \, .
\end{split}
\ee
The beta functions for the massless case follow straightforwardly by setting $m=0$. In this case the beta functions form an autonomous system of differential equations, independent of $\mu$. 

The system of equations \eqref{betamass} is the central result of this subsection. Owed to the fact that it captures the quantum effects arising from integrating out a minimally coupled scalar field, the system of flow equations \eqref{betamass} is structurally relatively simple. The beta function for the dimensionless Newton's constant is independent of all other couplings and can thus be solved on its own. Subsequently, this solution can be plugged into the remaining equations which can again be solved individually. Thus considering the ``healthy extension'' does not modify the renormalization group flow in the ``projectable'' sector of the theory. In this sense the flows found for the ``healthy extension'' form a direct extension of the ones obtained in the projectable theory.

We close this subsection with the following remark. Albeit the full set of heat-kernel coefficients for the relevant and marginal interaction terms of the ``healthy extension'' is not known, it is clear that the beta functions
for the couplings $\ut_{n,i}$ will have the same structure as the ones for the couplings $\gt_{n,i}$ with different coefficients $b_n a_{2n,i}$. Parameterizing the unknown coefficients by $c_{n,i}$ where $2n$ is again
the number of spatial derivatives in the interaction monomials and $i$ enumerates the corresponding basis monomials, one concludes that
\be\label{healthyext}
	\b_{\ut_{n,i}} =   -2 \, \xi_n \, \ut_{n,i} + g \, (4\pi)^{-(d-1)/2} \,   e^{-\m^{-2}m^2} \left(4 \, c_{n,i} - \tfrac{2}{3} \, \ut_{n,i} \, e_2 \right) \, . \\
\ee
As we will see in the next subsection, this parametric form of the beta functions already allows to draw conclusions about the structure of the renormalization group flow.  

%-------------------------------------------------------------
\subsection{Renormalization group flows at criticality}
%-------------------------------------------------------------
In terms of HL gravity, the most interesting situation is when
the system is at criticality with $d=z=3$. In this case, Newton's constant is marginal and the theory is power-counting renormalizable. We will discuss the renormalization group flows entailed by \eqref{betamass} and the extension \eqref{healthyext} in this subsection. \\

\noindent
{\bf Massless flows} \\
Restoring the number of scalar fields $n_s$, the scalar-induced massless beta functions for the kinetic terms at criticality simplify to
\be\label{critkin}
\beta_g = \frac{n_s}{5 \pi} \, g^2 \, , \qquad \beta_\lambda = \frac{n_s}{15\pi} \left(3 \lambda - 1 \right) \, g \, . 
\ee
This system is supplemented by the beta functions for the potential couplings in the projectable sector
\be\label{critproj}
\begin{split}
	\b_{\gt_{0}} =  & \, - 2 \, \gt_{0} + n_s \,  \tfrac{g}{\pi} \,    \left( \, b_0  + \tfrac{1}{5} \, \gt_{0} \right) \, \\
	\b_{\gt_{1}} =  & \, - \tfrac{4}{3} \, \gt_{1} + n_s \,  \tfrac{g}{\pi} \,    \left( \, b_1 \, a_{2,i}  + \tfrac{1}{5} \, \gt_{1} \right) \, \\
	\b_{\gt_{2,i}} = & \,  - \tfrac{2}{3} \, \gt_{2,i} + n_s \, \tfrac{g}{\pi}  \left(\, b_2 \, a_{4,i} + \tfrac{1}{5} \, \gt_{2,i}  \right) \, , \\
	\b_{\gt_{3,i}} = & \,  n_s \, \tfrac{g}{\pi}  \left(\, b_3 \, a_{6,i} + \tfrac{1}{5} \, \gt_{3,i}  \right) \, ,
\end{split}
\ee
and the ``healthy extension''
\be\label{crithealthy}
\begin{split}
	\b_{\ut_{1}} =  & \, - \tfrac{4}{3} \, \ut_{1} + n_s \,  \tfrac{g}{\pi} \,    \left( \, c_1 + \tfrac{1}{5} \, \ut_{1} \right) \, , \\
		\b_{\ut_{2,i}} = & \,  - \tfrac{2}{3} \, \ut_{2,i} + n_s \, \tfrac{g}{\pi}  \left(\, c_{2,i} + \tfrac{1}{5} \, \ut_{2,i}  \right) \, , \\
	\b_{\ut_{3,i}} = & \,  n_s \, \tfrac{g}{\pi}  \left(\, c_{3,i} + \tfrac{1}{5} \, \ut_{3,i}  \right) \, .
\end{split}
\ee
The beta functions \eqref{critkin} and the marginal couplings in the projectable sector agree with the ones obtained from the functional renormalization group  \cite{D'Odorico:2014iha}. This match is in agreement with the 
 expectation that the one-loop beta functions for marginal couplings are universal in the sense that they do not depend on the regularization scheme. Typically, this universality does not extent to relevant coupling constants. On this basis, it is not expected that the beta functions for $\gt_0$, $\gt_1$ and $\gt_{2,i}$ agree in the two computations.

\begin{figure}
	\begin{center}
		\includegraphics[scale=0.85]{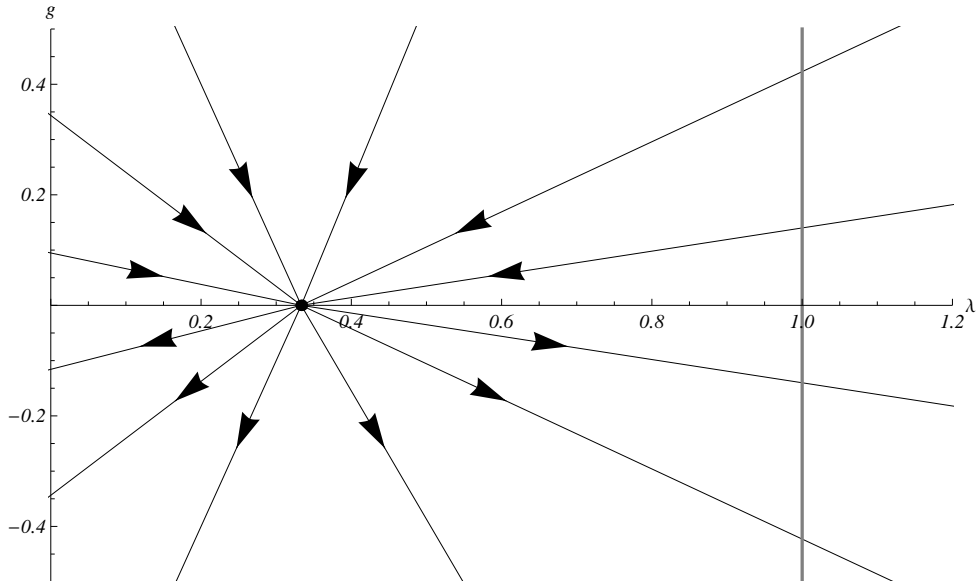}
		\caption{\label{attractor} RG flow in the $g$--$\lambda$ plane for the massless scalar case in $d=z=3$.
			The vertical line represents the isotropic plane $\l=1$.
			The arrows point towards the IR, and the black dot indicates the position of the aGFP.
			We clearly see that the aGFP is an IR attractor for $g>0$ and a UV attractor for $g<0$.}
	\end{center}
\end{figure}
The most prominent feature of the system \eqref{critkin}, \eqref{critproj} and \eqref{crithealthy} is the presence of an anisotropic Gaussian fixed point (aGFP) where all beta functions vanish simultaneously
\be\label{aGFP}
\begin{split}
g^* = 0 \, , \quad \lambda^* = \frac{1}{3} \, , \quad
 & \, \gt^*_{0} = \gt^*_{1} = \gt^*_{2,i} = 0 \, , \quad 
\gt^*_{3,i} = -5 \, b_3 \, a_{6,i} \, ,  \\ 
& \, \ut^*_{1} = \ut^*_{2,i} = 0 \, , \qquad \quad
\ut^*_{3,i} = -5 \, c_{3,i} \, .
\end{split}
\ee
This fixed point is two-fold degenerate in the sense that it is
located at the intersection of a line of Gaussian fixed points with $g=0$ and a line of non-Gaussian fixed points parameterized by the anisotropy parameter $z$ \cite{D'Odorico:2014iha}. The aGFP corresponds to a free theory where the propagators exhibit anisotropic scaling with $z=3$.  Notably, the position of the fixed point is independent of $n_s$. This implies that the fixed point solution becomes exact in the large-$n_s$ limit, where the contribution of the gravitational loops to the beta functions is negligible. It is actually this fixed point which underlies the expected renormalizability of HL gravity both in the projectable as well as in the non-projectable version.

At this stage, it is instructive to investigate the subsystem \eqref{critkin} in order to understand the properties of the aGFP in more detail. 
Fig.\ \ref{attractor} shows a set of sample solutions illustrating the renormalization group flow in the $g$--$\lambda$ plane with the aGFP marked by a black dot. This establishes that the aGFP actually acts as an IR attractor for trajectories starting with a positive Newton's constant. The analytic solution of the system \eqref{critkin} is 
\be
g(\mu) = - \frac{5 \pi}{5 \pi a_1 + \log \mu} \, , \quad \lambda(\mu) = \frac{a_2}{5 \pi a_1 + \log\mu} + \frac{\log\mu}{3 (5\pi a_1 + \log\mu)} \, ,
\ee
where $a_1$ and $a_2$ are integration constants. Thus $g$ increases with increasing energy scale, and the system becomes strongly coupled at high energy. In fact, the dimensionless Newton's constant apparently becomes infinite at some finite energy, resulting in a Landau pole. In this respect, the renormalization group flow of the system resembles the features known from QED. \\

\begin{figure}
	\begin{center}
		\includegraphics[scale=0.8]{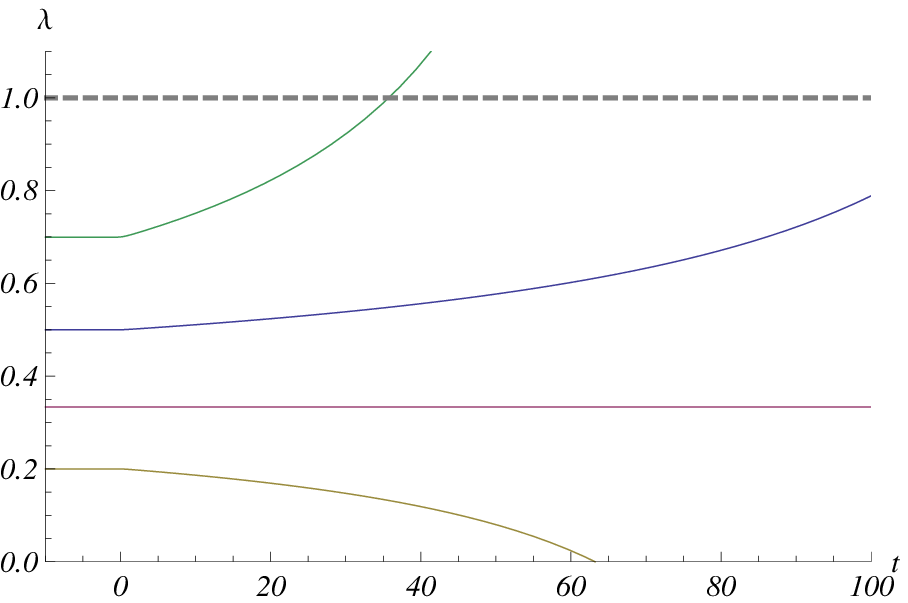}
		\includegraphics[scale=0.8]{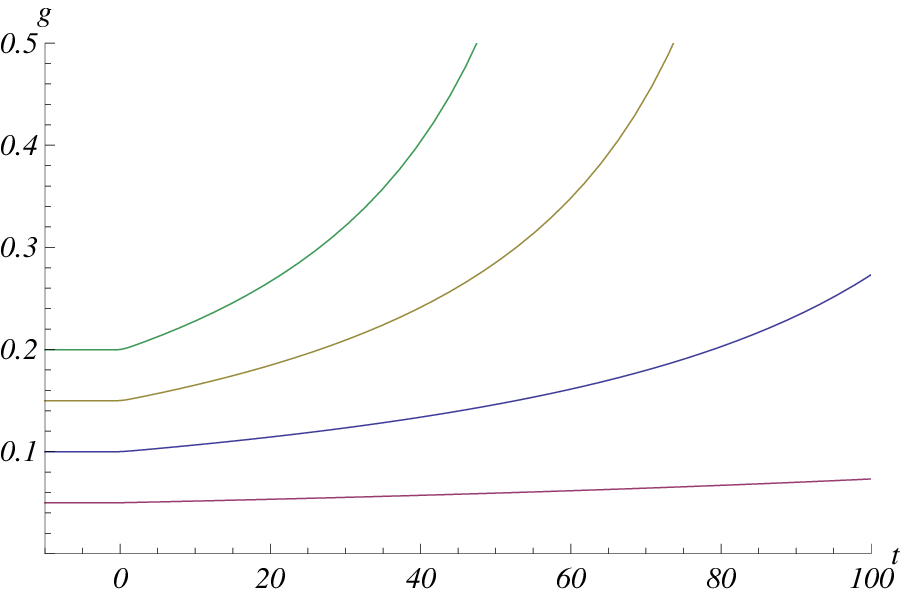}
		\caption{\label{tmassive} RG flow for $g$ and $\l$ as a function of $t = \ln(\mu/m)$. 
			The dashed line represents the isotropic plane $\l=1$.
			%The decoupling is manifest when the logarithm becomes negative.
			The flow of $g$ and $\l$ freezes out for $\mu^2 \lesssim m^2$.}
	\end{center}
\end{figure}

{\noindent}
{\bf Massive flows} \\
The beta functions incorporating a non-zero mass can be obtained from the system \eqref{critkin}, \eqref{critproj}, and \eqref{crithealthy} through the formal substitution $n_s \mapsto n_s \, e^{-m^2/\m^2}$. The presence of the mass scale motivates introducing the ``renormalization group time'' 
$t = \ln(\mu/m)$. Fig.\ \ref{tmassive} then illustrates the typical renormalization group flow of $g(t)$ and $\l (t)$. 
In this case the flow of the coupling constants freezes for $\mu \lesssim m$.
This is the standard decoupling phenomenon of a gapped theory. 
Due to the decoupling of the scalars below the scale $m^2$, the IR limit of the matter-induced flow is trivial and the infrared value of the coupling constants can be tuned to any value by choosing suitable initial conditions. An interesting observation is that for energies $\mu > m$,
the flow is still attracted towards the aGFP before the couplings freeze out.
This means that if the isotropic system is perturbed by a Lorentz breaking (scalar) operator at some high energy, the renormalization group flow will drive the system away from the isotropic plane. It would be very interesting to understand the phenomenological consequences of this instability in more detail.

%-------------------------------------------------------------
\section{Conclusions}
%-------------------------------------------------------------
One of the primary motivations for HL gravity is the prospect of obtaining a quantum theory for gravity which is both perturbatively renormalizable and unitary. Its core feature is the anisotropic scaling of spatial and time directions \eqref{scalingrel} which, at criticality, leads to a dimensionless Newton's constant without introducing more than two time derivatives in the fundamental action. This renders the theory power-counting renormalizable without introducing negative norm states plaguing higher-derivative gravity.  
An open challenge in the program consists of going beyond the power-counting arguments and establishing that HL gravity indeed possesses a suitable renormalization group fixed point which renders the theory asymptotically free. 

In this paper we have devised
a general method to derive renormalization group flows in HL gravity based on the effective action. Following a strategy developed for quantum field theory in curved spacetime (see, e.g., \cite{Buchbinder:1992rb,Avramidi:2000bm}), we used the Seeley-deWitt expansion of the heat kernel for anisotropic differential operators to compute the one-loop effective action originating from integrating out a (massive) scalar field with anisotropic dispersion relation minimally coupled to HL gravity. Our approach naturally links
with the background field method in curved space and directly leads to
a covariant effective action. Our technique is very general and can systematically be generalized to other types of matter fields including  fermions and gauge fields as well as to graviton loops. 

The main technical achievement of our work is the first systematic computation of the Seeley-deWitt expansion for an anisotropic Laplace-type operator $D^2 = \Delta_t + (\Delta_x)^z$ in a curved spacetime. The heat-kernel coefficients for the two-derivative terms and all intrinsic curvature monomials are obtained explicitly for arbitrary dimension of the spatial slices $d$ and anisotropy $z$. Our
derivation is based on off-diagonal heat-kernel techniques. Notably, the applicability of this method is not limited to a specific version of HL gravity: we explicitly demonstrated that the heat-kernel coefficients associated with operators appearing in the potential of the ``healthy extension'' can be obtained as well, even though their computation
is much more demanding than finding the corresponding coefficients appearing
in the projectable version of the theory. In fact, obtaining the complete set of heat-kernel coefficients associated with the power-counting relevant and marginal interaction terms in the ``healthy extension'' may be as complex
as a two-loop computation in general relativity and will thus requires a dedicated effort. We hope to come back to this point in the future.

The key features of the scalar-induced renormalization group flow at criticality $z=d=3$ are summarized as follows. In the case where the scalar field is massless, the flow exhibits an anisotropic Gaussian fixed point (aGFP) \cite{D'Odorico:2014iha}. For the relevant couplings of the theory, this fixed point acts as a UV attractor, while it is UV repulsive for a (dimensionless) positive Newton's constant. Tracing the renormalization group flow emanating from the aGFP in the IR towards the UV, the renormalization group trajectories terminate in a Landau pole. In the case where the scalar field has non-zero mass $m$ the renormalization group flow freezes at scales $\mu^2 \lesssim m^2$ and the value of the coupling constants in the IR constitute free parameters of the theory. Notably, these features are independent of the version of HL gravity under consideration and apply to the projectable and non-projectable 
versions of the theory.
This universality is easily traced back to the observation that the different versions of HL gravity differ in their gravitational sectors only. Since the present model does not include the contribution of graviton loops to the beta functions, our results are independent of the precise structure of these sectors.

The gravitational renormalization group flow induced by the massless scalar field is strikingly similar to the one encountered in QED. In both cases the fixed point corresponding to the free theory is an IR attractor and the high-energy behavior of the theory is screened by a Landau pole. Notably, this behavior becomes exact in the large $N$-limit when the number of scalar matter fields dominates over the gravitational contribution to the beta functions.
This does not imply, however, that HL gravity cannot be asymptotically free. Similarly to the one-loop beta function of QCD with $n_f$ flavors,
\be
\beta(g) = - \left( 11 - \frac{2 \, n_f}{3} \right) \, \frac{g^3}{16 \pi^2} \, ,
\ee
the sign of the beta function in HL gravity may depend on the type and precise number of matter fields appearing in the theory. It would be very interesting to determine this dependence in detail.

\vfill

\subsection*{Acknowledgements}
We thank J.\ Alexandre, J.\ Ambj{\o}rn, S.\ Gryb, R.\ Loll, E.\ van Loon, L.\ Pires, and M.\ Reuter for helpful discussions. The work of F.S.\ and G.D.\ is supported by the Netherlands Organization for Scientific Research (NWO) 
within the Foundation for Fundamental Research on Matter (FOM), grants 13PR3137 and 13VP12.

\pagebreak

%-------------------------------------------------------------
\begin{appendix}
%-------------------------------------------------------------

%-------------------------------------------------------------
\section{Off-diagonal heat-kernel techniques}
\label{App.A}
%-------------------------------------------------------------
The computation of the Seeley-DeWitt expansion for the operator $D^2$
crucially hinges on the ability to evaluate operator traces of
the form
\be\label{tr:canform}
\Tr_i \left[ \, e^{-s \Delta^{(D)}} \,  \cO \, \right] =  \langle x_i | \,  \e^{-s \Delta^{(D)}} \,  \cO \,| x_i \rangle \,
\ee
where $\Delta^{(D)} \equiv - g^{\mu\nu}D_\mu D_\nu$ is the standard second order Laplacian on a $D=d+1$-dimensional manifold, $\cO$ is an operator insertion typically including covariant derivatives $D_\mu$ constructed from $g_{\mu\nu}$, and the $| x_i \rangle$ form a basis of position eigenstates carrying internal indices $i$. In the present computation, the Laplacian acts on scalar fields so that the internal indices are absent.  

The off-diagonal heat-kernel allows to evaluate traces of the form
\eqref{tr:canform} as follows \cite{Anselmi:2007eq,Benedetti:2010nr}.\footnote{For some early work on the spin-dependence of the off-diagonal heat-kernel also see  \cite{Christensen:1978yd}.} Inserting a complete set of states,
eq.\ \eqref{tr:canform} can be written as 
\be\label{off-diag.trace}
\begin{split}
\langle x | \, \cO \, \e^{-s \Delta^{(D)}}  | x \rangle = & \, \langle x | \, \cO \, |x^\prime \rangle \langle x^\prime| \e^{-s \Delta^{(D)}}  | x \rangle \\
= & \,  \int d^Dx \sqrt{g} \, {\rm tr} \left[ \cO H(x, x^\prime; s) \right]_{x = x^\prime} \, ,
\end{split}
\ee
where $H(x, x^\prime; s) \equiv \langle x^\prime| \e^{-s \Delta^{(D)}}  | x \rangle$ is the heat kernel at non-coincident points. This ``off-diagonal'' heat-kernel obeys the asymptotic expansion
\be\label{eqa3}
H(x, x^\prime; s) \simeq  (4 \pi s)^{-D/2} \, \e^{- \frac{\sigma(x,x^\prime)}{2 s}} \, \sum_{n=0}^\infty s^n A_{2n}(x, x^\prime) \, .
\ee
Here, $\sigma(x,x^\prime)$ denotes half the squared geodesic distance between $x$ and $x^\prime$ satisfying
\be\label{eqa4}
\half \sigma^{;\mu} \sigma_{;\mu} = \sigma \, , \qquad \sigma(x,x) = \sigma_{;\mu}(x,x) = 0 \, , \qquad \sigma_{;\mu\nu}(x,x) = g_{\mu\nu}(x) \, , 
\ee
 and $A_{2n}(x, x^\prime)$, $n \in \mathbb{N}$, are the off-diagonal heat-kernel coefficients. The latter are  purely geometrical two-point objects, formally independent of the dimension $D$. At coinciding points $A_{2n}(x,x)$ agrees with the standard heat-kernel coefficients \eqref{eqb44}. Their values at non-coincident points can be encoded in a Taylor expansion around $x$ with coefficients given in \cite{Decanini:2005gt} and references therein.

In practice, eq.\ \eqref{off-diag.trace} is evaluated by acting with the covariant derivatives contained in $\cO$ on $H(x, x^\prime; s)$ and subsequently
taking the coincidence limit. In order to simplify the occurring structures
one can assume that the covariant derivatives within $\cO$ are symmetrized, i.e.,
\be\label{symop}
\cO = \sum_{k = 0}^n M^{\a_1 \ldots \a_{2k}} \, D_{(\a_1} \cdots D_{\a_{2k})} \, ,
\ee
with totally symmetric matrices $M^{\a_1 \ldots \a_{2k}}$. This form can always be achieved by successively writing non-symmetric combinations of the derivatives as a sum of symmetric and antisymmetric pieces and expressing the latter in terms of curvature tensors using commutator relations of the form
\be
\left[\, D_\mu \, , \, D_\nu \, \right] \, a_\lambda = R_{\mu\nu\lambda}{}^\sigma \, a_\sigma \, . 
\ee

Based on the structure \eqref{symop}, it then suffices to have an explicit
form for the quantities
\be\label{hlim}
H_{\a_1 \ldots \a_{2n}} \equiv \overline{D_{(\a_1} \cdots D_{\a_{2n})} \, H(x, x^\prime; s) } \, ,
\ee
where the overline denotes the coincidence limit $x \rightarrow x^\prime$ and the round brackets indicate the symmetrization of indices with unit strength, $(\a\b) = \half(\a\b +\b\a)$. In terms of the $H$-tensors
the trace \eqref{tr:canform} takes the rather simple form
\be\label{trace:inv}
\Tr \left[ \, e^{-s \Delta^{(D)}} \,  \cO \, \right] = 
\int d^Dx \sqrt{g} \,   \sum_{k \ge 0} \, M^{\a_1 \ldots \a_{2k}} \, H_{\a_1 \ldots \a_{2k}} \, . 
\ee

The explicit expressions for the $H$-tensors can be constructed
from the covariant derivatives of $\sigma(x, x^\prime)$ and $A_{2n}(x,x^\prime)$ 
evaluated in the coincidence limit.
Based on the properties \eqref{eqa4} one can prove recursively that $\sigma(x,x^\prime)$ satisfies \cite{Anselmi:2007eq}
\be\label{sigmarel}
\overline{\sigma(x, x^\prime)_{;\mu(\alpha_1 \ldots \alpha_n)}} = 0 \, , \qquad n \ge 2 \, .
\ee
Furthermore, a recursion relation based on the heat equation gives \cite{Anselmi:2007eq}
\be\label{arel}
\renewcommand{\arraystretch}{1.5}
\begin{array}{lll}
\overline{A_0(x, x^\prime)} = 1 \, , \quad  & \overline{A_{0}(x, x^\prime)_{;\mu}} = 0 \, , \quad & \overline{A_{0}(x, x^\prime)_{;(\mu\nu)}} = \frac{1}{6} R_{\mu\nu} \, , \\ \overline{A_2(x, x^\prime)} = \frac{1}{6} R \, .
\end{array}
\ee
Coefficients $A_{2n}(x,x^\prime)$ with $n \ge 2$ and terms based on $A_{0}(x, x^\prime)$ and $A_{2}(x, x^\prime)$ involving additional covariant derivatives
contain more than two covariant derivatives and do not contribute to the computations carried out in this work. They can be found in \cite{Groh:2011dw}. Taking covariant derivatives of \eqref{eqa3}, evoking the coincidence limit, and substituting the relations \eqref{eqa4}, \eqref{sigmarel} and \eqref{arel} yields the following asymptotic expansion of the $H$-tensors 
\be\label{atexplicit}
\begin{split}
	H \simeq & \, \frac{1}{(4\pi s)^{D/2}} \left(1 + \tfrac{1}{6} s R \right) \, , \\
H_\alpha \simeq & \, 0 \, , \\	
H_{\a\b} \simeq & \, \frac{1}{(4\pi s)^{D/2}} \Big\{ -  \frac{1}{2s} \, g_{\a\b} \,  \left(1 + \tfrac{1}{6} s R \right) + \frac{1}{6} R_{\a\b} \Big\} \, , \\
H_{\a\b\mu} \simeq & \, 0 \, , \\
H_{\a\b\m\n} \simeq & \, \frac{1}{(4\pi s)^{D/2}} \Big\{ \frac{1}{4s^2} \left( g_{\a\m} g_{\b\n} + g_{\a\n} g_{\b\m} + g_{\a\b} g_{\m\n} \right) \left(1 + \tfrac{1}{6} s R \right) \\
&  - \frac{1}{12s} \left( g_{\a\m} R_{\b\n} + g_{\a\n} R_{\b\m} + g_{\b\m} R_{\a\n} + g_{\b\n} R_{\a\m} + g_{\a\b} R_{\m\n} + g_{\m\n} R_{\a\b} \right)
\Big\} \, ,
\end{split}
\ee
which is exact up to terms including three or more derivatives.
Eq.\ \eqref{eqa3} in combination with the $H$-tensors \eqref{atexplicit}
constitutes the central result of this appendix. In App.\ \ref{App.B} we use
these formulas to evaluate the operator traces encoding the Seeley-deWitt coefficients of $D^2$.

%-------------------------------------------------------------
\section{Anisotropic heat-kernel coefficients: derivation}
\label{App.B}
%-------------------------------------------------------------
This appendix provides the technical details
underlying the derivation of the heat-kernel coefficients
for the anisotropic differential operator \eqref{diffop}
reported in Sect.\ \ref{sect.3}.

%-------------------------------------------------------------
\subsection{General setup}
%-------------------------------------------------------------
The goal is to determine the coefficients
appearing in the Seeley-deWitt expansion \eqref{mastermaster}
of
\be\label{optrace}
K_{D^2}(s) = {\rm Tr} \, e^{-sD^2}
\ee
for the anisotropic differential operator $D^2 = \Delta_t + (\Delta_x)^z$
defined in eq.\ \eqref{diffop}. A key advantage of the Seeley-deWitt expansion of $D^2$ is its covariance with respect to the symmetry group \eqref{fpdiff}.
The structure of the interaction monomials is then largely determined
by these symmetries. This feature allows to simplify the
computation of the heat-kernel coefficients
by working with a specific subclass of metrics in order to simplify the  computational effort. This class
has to be sufficiently general in order to uniquely identify
the curvature monomials of interest. Covariance of the heat-kernel
then implies that the coefficients determined from this specific set of backgrounds carry over to the general case.

 In practice the heat-kernel coefficients of $D^2$
will be obtained by recasting the operator trace \eqref{optrace}
into the form
\be\label{comtrace}
{\rm Tr} \, e^{-sD^2} = {\rm Tr} \left[ \, e^{-s \Delta^{(D)}} \, \cO \, \right] \, , 
\ee
where $\Delta^{(D)} = D^2|_{z=1}$ is the standard Laplace operator \eqref{Dlaplacian} and the operator insertions $\cO$ are built from commutators of $\Delta_t$ and $\Delta_x$. The rhs of this expression can be evaluated
by employing the off-diagonal heat-kernel techniques described in App.\ \ref{App.A}. The relation \eqref{comtrace} is systematic in the sense that at a given order of the Seeley-deWitt expansion multi-commutators contribute up to a finite order only.\footnote{In the sequel, we restrict ourselves to multi-commutators with at most two spatial or time derivatives acting on the component fields. Commutators containing more derivatives of the component fields do not contribute to the heat-kernel coefficients computed in this paper and will not be considered.}

The first step towards obtaining \eqref{comtrace} uses the 
Zassenhaus or inverse Campbell-Baker-Hausdorff formula
\begin{equation}\label{CBHformula}
e^{-s(A+B)}=e^{-sA}e^{-sB}e^{-\frac{s^2}{2}[A,B]}e^{-\frac{s^3}{6}([A,[A,B]]+2[B,[A,B]])}\cdots
\end{equation}
to split the exponential
\begin{equation}\label{eqb4}
{\rm Tr} \, e^{-sD^2} ={\rm Tr}\left[ e^{-s\Delta_t} e^{-s(\Delta_x)^z} \, {\cal C}(\Delta_t,\Delta_x)\right] \, . 
\end{equation}
The operator insertion $\cC$ has an analytic expansion in $s=0$ whose coefficients can be found by expanding the commutator terms in eq.\ \eqref{CBHformula}. Retaining all commutators which contain up to 
two time derivatives acting on the component fields
\be\label{eqB5}
\begin{split}
{\cal C}(\Delta_t,\Delta_x) \simeq & \, 1 -\frac{s^2}{2}[\Delta_t,(\Delta_x)^z] -\frac{s^3}{6}([\Delta_t,[\Delta_t,(\Delta_x)^z]]+2[(\Delta_x)^z,[\Delta_t,(\Delta_x)^z]])  \\
& \, +\frac{s^4}{8}[\Delta_t,(\Delta_x)^z]^2 + \cO(s^5) \, . 
\end{split}
\ee
Generically, these commutators involve non-integer powers of $\Delta_x$. By induction one can prove that these are
related to the standard commutators of $\Delta_t$ and $\Delta_x$ by
\be\label{eqB6}
\left[ \, \Delta_t \, , \, (\Delta_x)^z \, \right] = z \, (\Delta_x)^{z-1} \, \left[ \, \Delta_t \, , \, \Delta_x \, \right] + 
\half z(z-1) (\Delta_x)^{z-2} \left[ \left[ \, \Delta_t \, , \, \Delta_x \, \right], \Delta_x \right] + \ldots \, , 
\ee
where the dots contain multi-commutators of higher order. Without loss of generality, we assume that all terms in $\cC(\Delta_t,\Delta_x)$ are ``normal ordered'' in the sense that all operators $\Delta_x$ are at the left and the $\Delta_t$ appear at the right,
\be
\cC(\Delta_t,\Delta_x) = \sum_i \, f_{x,i}(\Delta_x) \, \tilde{\cC}_i \, f_{t,i}(\Delta_t) \, . 
\ee
The $\tilde{\cC}_i$ contain all parts of the commutators which do not combine into the Laplacians. Using the cyclicity of the trace, eq.\ \eqref{eqb4} is cast into the form
\begin{equation}\label{eqb12}
{\rm Tr} \, e^{-sD^2} = \sum_i {\rm Tr}\left[ f_{t,i}(\Delta_t) \, e^{-s\Delta_t} \, e^{-s(\Delta_x)^z} \, f_{x,i}(\Delta_x)  \,\tilde{{\cal C}} \, \right] \, . 
\end{equation}

The goal of the next step is to eliminate the exponent $z$ appearing in $e^{-s(\Delta_x)^z}$. For this purpose, we introduce the anti-Laplace transform $\widetilde{W}(v)$ of a function $W(x)$ as
\be\label{eqb9}
W(x)  \equiv \int_0^\infty dv \, \widetilde{W}(v) \, e^{-xv} \, . 
\ee
Setting
\be\label{Wfct}
 W_i(x) = e^{-sx^z} \, f_{x,i}(x)
\ee
one arrives at
\be\label{eqB10}
{\rm Tr} \, e^{-sD^2} = \sum_i \int_0^\infty dv \, \widetilde{W}_i(v) \, {\rm Tr}\left[ f_{t,i}(\Delta_t) \, e^{-s\Delta_t} \, e^{-v\Delta_x} \,   \,\tilde{{\cal C}} \, \right] \, . 
\ee
In principle, the functions $f_{t,i}$ may also be eliminated by a suitable Laplace transform. Since $f_{t,i} = (\Delta_t)^{n_i}$ with $n_i$ being integer, it is more convenient to generate these terms by taking suitable derivatives
of the trace with respect to $s$
\be\label{eqB11}
{\rm Tr} \, e^{-sD^2} = \sum_i \int_0^\infty dv \, \widetilde{W}_i(v) \, \left( -\frac{\partial}{\partial s} \right)^{n_i} {\rm Tr}\left[ \, e^{-s\Delta_t} \, e^{-v\Delta_x} \,   \,\tilde{{\cal C}} \, \right] \, . 
\ee

In the next step, one introduces an auxiliary Laplace operator
\be
\tilde{\Delta}_x = \frac{v}{s} \, \Delta_x \, .
\ee
This auxiliary operator is associated with the auxiliary spatial metric
\be\label{auxspatmet}
 \tilde{\sigma}_{ij}=\frac{s}{v}\, \sigma_{ij} \, . 
\ee
Writing \eqref{eqB11} in terms of this auxiliary Laplacian allows to combine the two exponentials by applying the Campbell-Baker-Hausdorff formula in 
	its standard form
\begin{equation}\label{eqb15}
e^{-s \Delta_t} e^{-s \tilde{\Delta}_x}=e^{-s (\Delta_t +\tilde{\Delta}_x)}{\cal B}(\Delta_t, \tilde{\Delta}_x)
\end{equation}	
with
\begin{equation}\label{eqb16}
\begin{split}
{\cal B}(\Delta_t, \tilde{\Delta}_x) \simeq  & \, 1 +\frac{s^2}{2} [\Delta_t, \tilde{\Delta}_x] +\frac{s^3}{6} \left( [\Delta_t,[\Delta_t, \tilde{\Delta}_x]] +2[\tilde{\Delta}_x,[\Delta_t, \tilde{\Delta}_x]] \right) \\ & \, +\frac{s^4}{8} [\Delta_t, \tilde{\Delta}_x]^2 + {\cal O}(s^5) \, . 
\end{split}
\end{equation}
Substituting this result into \eqref{eqB11}, the rhs indeed assumes
the form \eqref{comtrace} with the $D$-dimensional Laplacian $\tilde \Delta^{(D)} \equiv \Delta_t + \tilde{\Delta}_x$ defined with 
respect to the auxiliary spatial metric \eqref{auxspatmet}
\be
{\rm Tr} \, e^{-sD^2} = \sum_i \int_0^\infty dv \, \widetilde{W}_i(v) \, \left( -\frac{\partial}{\partial s} \right)^{n_i} {\rm Tr}\left[ \, e^{-s\tilde{\Delta}^{(D)}} \, \cB  \,\tilde{{\cal C}} \, \right] \, .
\ee
This operator trace is then readily evaluated by applying the off-diagonal heat-kernel techniques of App.\ \ref{App.A}.  Expanding all curvatures into $d$--dimensional ADM invariants by using the Gauss-Codazzi relations and 
converting back to the original metric $\sigma_{ij}$ the rhs is 
readily expressed in terms of curvature monomials.

The final $v$-integral can be evaluated using Mellin-transform
techniques, see, e.g.\ \cite{Reuter:1996cp,Codello:2008vh}. Using
\be\label{eqb18}
\begin{split}
Q_n[W] \equiv & \, \int_0^\infty dv \, v^{-n} \, \widetilde{W}(v)  \, , \quad \qquad \qquad n \ge 0 \\
= & \, \frac{1}{\Gamma(n)} \, \int_0^\infty dx \, x^{n-1} \, W(x) \, , 
\end{split}
\ee
the integrals over the anti-Laplace transform $\widetilde{W}(v)$ are
related to moments of the initial function $W(x)$. Combining
eqs.\ \eqref{eqB5} and \eqref{eqB6}, one finds that the functions
$W_i(x)$, eq.\ \eqref{Wfct}, fall into the classes collected in Tab.\ \ref{integrals}. Their $Q$-functionals
\begin{table}[t]
	\renewcommand{\arraystretch}{1.5}
	\begin{center}
		\begin{tabular}{ll}
			\hline\hline
			$W_1(x) \equiv \, e^{-s x^z} \, , $ & 
			$ \int_0^\infty dv \, v^{-d/2} \, \widetilde{W}_1(v) = 
			s^{-d/(2z)} \, \frac{\Gamma(d/(2z))}{z \, \Gamma(d/2)}$\\
			$W_2(x) \equiv  \, s \, z \, x^{z-1} \, e^{-s x^z} \, , $ & 
			$ \int_0^\infty dv \, v^{-(d/2+1)} \, \widetilde{W}_2(v) = s^{-d/(2z)} \, \frac{\Gamma(d/(2z))}{z \, \Gamma(d/2)} $ \\
			$W_3(x) \equiv  \, s^2 \, z^2 \, x^{2z-2} \, e^{-s x^z} \, , $ & 
			$\int_0^\infty dv \, v^{-(d/2+2)} \, \widetilde{W}_3(v) = s^{-d/(2z)} \, \frac{\Gamma(d/(2z))}{z \, \Gamma(d/2)} \, \frac{d+2z}{d+2} $ \\
			$W_4(x) \equiv \, \half \, s \, z (z-1) \, x^{z-2} \, e^{-s x^z} \, , $  \quad \quad &
			$\int_0^\infty dv \, v^{-(d/2+2)} \, \widetilde{W}_4(v) =  s^{-d/(2z)} \, \frac{\Gamma(d/(2z))}{z \, \Gamma(d/2)} \, \frac{z-1}{d+2} $ \\ 
			\hline\hline
		\end{tabular}
		\caption{\label{integrals} Integrals occurring in the evaluation of the traces \eqref{eqb25} for the extrinsic curvatures and monomials of the healthy extension. The functions $\widetilde{W}_i$ are related to the functions $W_i$ via the integral transform \eqref{eqb9}.}
	\end{center}
\end{table}
coincide with integral representations of the Gamma-function, which allows  determining the heat-kernel coefficients in closed form.
	   
%------------------------------------------------------
\subsection{Kinetic (extrinsic) terms}
\label{sect.B3}
%------------------------------------------------------
In order to obtain the heat-kernel coefficients built from
the extrinsic curvature \eqref{extcurve} it is convenient 
to choose a class of backgrounds $\cM = S^1 \times_t T^d$
where the radii of the tori depend on $t$. Explicitly,
we set
\begin{equation}\label{back1}
N=1 \, , \quad N^i=0 \, , \quad \sigma_{ij}(t,x)=\sigma_{ij}(t) \, .
\end{equation}
In this case the intrinsic curvatures and the derivatives of the lapse function vanish while the extrinsic curvature simplifies to $K_{ij} = \half \p_t \sigma_{ij}$. The operators \eqref{DtDx} take the form
\be\label{op1}
\Delta_t = - \p_t^2 - \sigma^{ij} K_{ij} \p_t \, , \quad \Delta_x = - \sigma^{ij} \, \p_i \p_j \, , 
\ee
while the non-vanishing Christoffel symbols of the the $D$-metric on $\cM$
are given by
\be\label{chris1}
\Gamma^t{}_{ij} = - K_{ij} \, , \quad \Gamma^i{}_{tj} = \sigma^{ik} K_{kj} \, . 
\ee
 Evaluating the trace \eqref{optrace} on the background \eqref{back1} projects the rhs of 
the master equation \eqref{mastermaster} onto the terms constructed from $K_{ij}$
\begin{equation}\label{eqb13}
\left. {\rm Tr} \, e^{-s D^2} \right|_{\cM} \simeq (4\pi)^{-(d+1)/2} \, s^{-\frac{1+d/z}{2}} \, \int  dt d^dx \sqrt{\sigma} \left[ b_0 + \tfrac{s}{6} e_1 K_{ij} K^{ij} + \tfrac{s}{6} e_2 K^2 + \ldots \right]
\end{equation}
where the dots indicate terms containing more than two time derivatives.
The aim is to determine the unknown coefficients $b_0$, $e_1$ and $e_2$.

This computation proceeds along the lines outlined in
the previous subsection. Starting from \eqref{op1} the commutators
entering into \eqref{eqB6} and \eqref{eqb16} are given by
\be\label{com1}
\begin{split}
\left[\Delta_t,\Delta_x \right] \phi =& \, -2 ( 2 K^{ij} \partial_t +\dot{K}^{ij} +K K^{ij} ) \partial_i \partial_j \phi \, , \\
\left[\Delta_t, \left[ \Delta_t,\Delta_x \right] \right] \phi = & \, 8 \dot{K}^{ij} \partial_t^2 \partial_i \partial_j \phi \, , \\
\left[\Delta_x, \left[ \Delta_t,\Delta_x \right] \right] \phi  = & \, 8 K^{ij} K^{lk} \partial_i \partial_j \partial_l \partial_k \phi \, , 
\end{split}
\ee
where the dot indicates a partial derivative with respect to $t$ and all terms with more than two derivatives acting on 
component fields have been dropped.

The next step constructs the explicit form of \eqref{eqb4} by
substituting \eqref{eqB5}. Conveniently, the result
is decomposed according to 
\be\label{eqb25}
\begin{split}
{\rm Tr} \, e^{-s D^2} \simeq & \, T_1 + T_2 + T_3 + T_4 + T_5 \, , 
\end{split}
\ee
with one trace associated with each commutator
\be\label{Tcommutators}
\begin{split}
 T_1 \equiv & \, {\rm Tr} \left[ \, e^{-s \Delta_t} \, e^{-s (\Delta_x)^z} \, \right] \, , \\ 
 T_2 \equiv & \, - \frac{s^2}{2} \, {\rm Tr} \left[ \, e^{-s \Delta_t} \, 
 e^{-s (\Delta_x)^z} \, \left[ \Delta_t, (\Delta_x)^z \,  \right] \right]\, , \\
 T_3 \equiv & \,  \frac{s^4}{8} \, {\rm Tr} \left[ \, e^{-s \Delta_t} \, e^{-s (\Delta_x)^z} \, \left( \left[ \Delta_t, (\Delta_x)^z \,  \right] \right)^2 \right] \, , \\
 T_4 \equiv & \, - \frac{s^3}{3} \, {\rm Tr} \left[ \, e^{-s \Delta_t} \, e^{-s (\Delta_x)^z} \, \left[ \, (\Delta_x)^z \, , \, \left[ \, \Delta_t \, , \,   (\Delta_x)^z \right]\right] \,  \right] \, , \\
 T_5 \equiv & \, - \frac{s^3}{6} \, {\rm Tr} \left[ \, e^{-s \Delta_t} \, e^{-s (\Delta_x)^z} \, \left[ \, \Delta_t \, , \, \left[ \, \Delta_t \, , \,   (\Delta_x)^z \right]\right] \,  \right] \, .
\end{split}
\ee
The commutators appearing in these expressions are subsequently expressed
through the basic commutators \eqref{com1} via the relation \eqref{eqB6}. Investigating the resulting operator insertions, one finds that the traces $T_3$, $T_4$, and $T_5$ already contain the maximum number of two time derivatives acting on spatial metrics $\sigma_{ij}$. At this point
all operators can be commuted freely and the partial derivatives may be promoted to covariant derivatives $D_\mu$ with respect to the $D$-metric on $\cM$. The commutators and connection pieces include further powers of the extrinsic curvature and therefore do not contribute to the present computation. 

We illustrate these steps by the detailed computation of $T_3$. The  substitution of the explicit commutators gives
\be
T_3 \simeq  2 \, s^4 z^2 \, {\rm Tr} \left[ \, e^{-s \Delta_t} \, e^{-s (\Delta_x)^z} \, \left(\Delta_x\right)^{2z-2} \, K^{ij} K^{kl} \, \p_i \p_j \p_k \p_l \, \p_t^2 \right] \, .
\ee
Subsequently, the trace is written in terms of the Laplace anti-transform $\widetilde{W}_3$ of the function $W_3(x) = s^2 z^2 x^{2z-2} e^{-s x^z}$ and
 the partial derivatives are completed into covariant derivatives $D_\mu$
\be
T_3 \simeq - 2 s^2 \int_0^\infty dv \, \widetilde{W}_3(v) \,  {\rm Tr} \left[ \Delta_t \, e^{-s \Delta_t} \, e^{-v \Delta_x} \, K^{ij} K^{kl} \, D_i D_j D_k D_l \right] \, .
\ee
The operator insertion $\Delta_t$ is then converted into a derivative of the trace with respect to $s$ and the two exponentials combine into the auxiliary Laplacian $\tilde{\Delta}^{(D)}$ constructed from the rescaled spatial metric \eqref{auxspatmet}
\be
T_3 \simeq - 2 s^2 \int_0^\infty dv \, \widetilde{W}_3(v) \, \left(-\frac{\p}{\p s}\right) \,  {\rm Tr} \left[   e^{-s \tilde{\Delta}^{(D)}} \, K^{ij} K^{kl} \, D_i D_j D_k D_l \right] \, .
\ee
At this stage the trace can easily be evaluated using the off-diagonal heat-kernel. Substituting \eqref{atexplicit} with the curvature terms set to zero gives
\be
T_3 \simeq  2 s^2 \int_0^\infty dv \, \widetilde{W}_3(v) \, \partial_s \, \left( \tfrac{1}{(4\pi s)^\frac{d+1}{2}} \, \int dt d^dx \sqrt{\tilde{\sigma}} K^{ij} K^{kl} \tfrac{1}{4s^2} \left( \tilde{\sigma}_{ik} \tilde{\sigma}_{jl} + \tilde{\sigma}_{il} \tilde{\sigma}_{jk} + \tilde{\sigma}_{ij} \tilde{\sigma}_{kl}  \right) \right)  \, .
\ee
Converting back to the initial spatial metric $\sigma$ via \eqref{auxspatmet} and taking the $s$-derivative, one arrives at 
\be
T_3 \simeq - \tfrac{1}{4} \left(4\pi \right)^{-(d+1)/2} s^{\frac{1}{2}} \int_0^\infty dv \, v^{-(d/2+2)}\, \widetilde{W}_3(v) \,  \, \int dt d^dx \sqrt{\sigma} \left[ 2  K^{ij} K_{ij} + K^2 \right]  \, .
\ee
Finally, the integral is carried out with the help of Tab.\ \ref{integrals},
yielding the final form of the trace 
\be\label{T3trace}
 T_3 \simeq  \, - \frac{1}{4} \left(4\pi \right)^{-(d+1)/2}  s^{-\tfrac{1}{2}(1+d/z)} \, \frac{\Gamma(\tfrac{d}{2z})}{z \, \Gamma(\tfrac{d}{2})} \, \frac{d+2z}{d+2}\, \int dt d^dx \sqrt{\sigma} \, s \, \left[ 2 K_{ij}K^{ij} + K^2 \right] . \\
\ee
Computing $T_4$ and $T_5$ proceeds along the same lines which results in
\be\label{T4trace}
\begin{split}
T_4 \simeq & \, - \frac{2}{3} \left(4\pi \right)^{-(d+1)/2}  s^{-\tfrac{1}{2}(1+d/z)} \, \frac{\Gamma(\tfrac{d}{2z})}{z \, \Gamma(\tfrac{d}{2})} \, \frac{d+2z}{d+2}\, \int dt d^dx \sqrt{\sigma} \, s \, \left[ 2 K_{ij}K^{ij} + K^2 \right] , \\
T_5 \simeq & \, \frac{1}{3} \left(4\pi \right)^{-(d+1)/2}  s^{-\tfrac{1}{2}(1+d/z)} \, \frac{\Gamma(\tfrac{d}{2z})}{z \, \Gamma(\tfrac{d}{2})} \, \frac{d+2z}{d+2} \, \int dt d^dx \sqrt{\sigma} \, s \, \left[ 2 K_{ij}K^{ij} + K^2 \right] . \\
\end{split}
\ee

The evaluation of $T_1$ and $T_2$ is slightly more involved
since there are additional contributions from the $\cB$-type 
commutators \eqref{eqb16} and the completion of the partial into covariant
derivatives. Upon applying \eqref{eqb15} the trace $T_1$ becomes
\begin{eqnarray}
T_1  \simeq && \int_0^\infty dv\, \widetilde{W}_1 \, {\rm Tr} \, e^{-s \tilde{\Delta}^{(D)}} 
\bigg( 1 +\frac{s^2}{2} [\Delta_t, \tilde{\Delta}_x] \nonumber \\ &&  +\frac{s^3}{6} \left( [\Delta_t,[\Delta_t, \tilde{\Delta}_x]] +2[\tilde{\Delta}_x,[\Delta_t, \tilde{\Delta}_x]] \right) +  
   \frac{s^4}{8} \left([\Delta_t, \tilde{\Delta}_x]\right)^2 \bigg)
\end{eqnarray}
with the function $W_1$ defined in Tab.\ \ref{integrals}. Substituting the
explicit commutators \eqref{com1} and applying the off-diagonal heat-kernel, the final result for $T_1$ is 
\be\label{T1trace}
\begin{split}
	T_1 \simeq  & \, \left(4\pi \right)^{-(d+1)/2}  s^{-\tfrac{1}{2}(1+d/z)} \, \frac{\Gamma(\tfrac{d}{2z})}{z \, \Gamma(\tfrac{d}{2})} \int dt d^dx \sqrt{\sigma} \, \left[1 + \tfrac{s}{6} \left( K^2 - K_{ij}K^{ij} \right) \right]  \\
	& + \, \tfrac{1}{12} \left(4\pi \right)^{-(d+1)/2}  s^{-\tfrac{1}{2}(1+d/z)} \, \frac{\Gamma(\tfrac{d}{2z})}{z \, \Gamma(\tfrac{d}{2})} \int dt d^dx \sqrt{\sigma} \, s \, \left[ 2 K_{ij}K^{ij} + K^2 \right] , \\
\end{split}
\ee
where the first and second line captures the contribution from the ``one'' and the commutator terms, respectively. Finally, the evaluation of $T_2$ yields
\be\label{T2trace}
\begin{split}	
	T_2 \simeq  & \,  \frac{1}{2} \left(4\pi \right)^{-(d+1)/2}  s^{-\tfrac{1}{2}(1+d/z)} \, \frac{\Gamma(\tfrac{d}{2z})}{z \, \Gamma(\tfrac{d}{2})} \, \frac{d+2z}{d+2}\, \int dt d^dx \sqrt{\sigma} \, s \, \left[ 2 K_{ij}K^{ij} + K^2 \right] \, . 
\end{split}
\ee

Summing up the contributions \eqref{T3trace}, \eqref{T4trace}, \eqref{T1trace} and \eqref{T2trace} gives the final result for the heat kernel evaluated on $\cM$
\be
\begin{split}
\left. {\rm Tr} \, e^{-s D^2} \right|_{\cM}  = & \, \left(4\pi \right)^{-(d+1)/2}  s^{-\tfrac{1}{2}(1+d/z)} \, \tfrac{\Gamma(\tfrac{d}{2z})}{z \, \Gamma(\tfrac{d}{2})} \int dt d^dx \sqrt{\sigma} \, \left[1 + \tfrac{s}{6} \left( K^2 - K_{ij}K^{ij} \right) \right]  \\ 
& - \,   \left(4\pi \right)^{-(d+1)/2}  s^{-\tfrac{1}{2}(1+d/z)} \, \tfrac{\Gamma(\tfrac{d}{2z})}{z \, \Gamma(\tfrac{d}{2})} \, \tfrac{z-1}{d+2} \, \int dt d^dx \sqrt{\sigma} \, \left[ \tfrac{s}{6} \left( K^2 + 2 K_{ij} K^{ij} \right) \right] \, . 
\end{split}
\ee
Comparing this result to the initial ansatz \eqref{eqb13} allows to read of the
coefficients 
\be\label{kincoeff}
b_0 = \frac{\Gamma(\tfrac{d}{2z})}{z \, \Gamma(\tfrac{d}{2})}
\, , \quad
e_1 = \frac{d-z+3}{d+2} \frac{\Gamma(\tfrac{d}{2z})}{z \Gamma(\tfrac{d}{2})} \, , \quad
e_2 = - \frac{d+2z}{d+2} \frac{\Gamma(\tfrac{d}{2z})}{z \Gamma(\tfrac{d}{2})} \, .
\ee
This result completes the derivation of the heat-kernel coefficients appearing 
in the kinetic sector of HL gravity.

%---------------------------------------------------------------
\subsection{Potential (intrinsic) terms}
\label{App:B3}
%---------------------------------------------------------------
In this subsection we determine the heat-kernel coefficients
associated with monomials constructed from the intrinsic
curvature. In this case it is convenient to evaluate the trace
\eqref{optrace} on the product space $\cM = S^1 \times \widetilde{\cM}$
where $\widetilde{\cM}$ is a $d$-dimensional compact manifold coinciding
with the spatial slices. The component fields of this background
are taken as
\be\label{back2}
N = 1 \; , \quad N^i = 0 \; , \quad \sigma_{ij}(t,x) = \sigma_{ij}(x) \, . 
\ee
so that all fields are independent of time. The operators \eqref{DtDx} in this background are given by
\be\label{op2}
\Delta_t = - \p_t^2 \, , \qquad \Delta_x = - \sigma^{ij} \nabla_i \nabla_j \, . 
\ee
where $\nabla_i$ is the covariant derivative constructed from $\sigma_{ij}$, i.e., $\nabla_k \sigma_{ij} = 0$.

 A direct consequence of the background \eqref{back2} is that the operators \eqref{op2} commute
\be\label{com2}
\left[ \, \Delta_t \, , \, \Delta_x \, \right] = 0 \, . 
\ee
The evaluation of the trace makes essential use of the product structure
\be\label{trprod}
\begin{split}
\left. {\rm Tr} \, e^{-s D^2} \right|_{\cM}  = & \,  {\rm Tr}\left[ e^{-s \Delta_t}\right] \, \cdot \, {\rm Tr}\left[ e^{-s (\Delta_x)^z}\right]  \\
= & \,  {\rm Tr}\left[ e^{-s \Delta_t}\right] \, \cdot \, \int_0^\infty dv \, \widetilde{W}_1(v) \, 	{\rm Tr}\left[ e^{-v \Delta_x }\right] \, . 
\end{split}
\ee
The first trace is the heat kernel on $S^1$,
\be\label{trS1}
 {\rm Tr}\left[ e^{-s \Delta_t}\right] = \frac{1}{(4 \pi s)^{1/2}} \, \int dt \, , 
\ee
while the second trace has the standard Seeley-deWitt expansion of the Laplacian on the $d$-dimensional manifold $\widetilde{\cM}$
\be\label{trMt}
{\rm Tr} \left[ e^{-v \Delta_x} \right] = \frac{1}{(4 \pi v)^{d/2}} \, \int d^dx \sqrt{\sigma} \, \sum_{n \ge 0} \, v^n \, a_{2n} \, . 
\ee
The coefficients $a_{2n}$ can readily be found in the literature
\cite{Vassilevich:2003xt,Gilkey:1995mj,Avramidi:2000bm,Groh:2011dw}.
In particular the first two are given by
\be\label{eqb44}
a_0 = 1 \, , \qquad a_2 = \tfrac{1}{6} R \, , \quad \ldots \, , 
\ee
with $R$ the (intrinsic) curvature on $\tilde{\cM}$. For manifolds of
dimension $d=2$ and $d=3$ where the curvature tensors obey special relations, the relevant coefficients $a_{2n}$ are listed in columns 3 and 6 of Tab.\ \ref{table}.

Substituting \eqref{trS1} and \eqref{trMt} into \eqref{trprod}
we find an expansion of the form
\be\label{heatpot}
\begin{split}
\left. {\rm Tr} \e^{-s D^2} \right|_{\cM} \simeq & \left(4\pi \right)^{-(d+1)/2}  s^{-\tfrac{1}{2}(1+d/z)} \, \sum_{n \ge 0 } \, \int dt d^dx \sqrt{\sigma}   \, s^{n/z} \,  b_n \, a_{2n} \, , 
\end{split}
\ee
where the coefficients $b_n$ are given by
\be\label{eqb46}
b_n(d,z) \equiv s^\frac{d-2n}{2z} \, \int_0^\infty dv \, v^{-d/2+n} \,  \widetilde{W}_1(v) \, . 
\ee
Here the $s$-dependent prefactor has been chosen in such a way to cancel the $s$-dependence of the integral, so that the coefficients $b_n(d,z)$ are independent of $s$. The explicit value of the integral may again be found
by applying the Mellin transform techniques \eqref{eqb18} which leads to the results \eqref{eq34} and \eqref{bngreaterdover2}.

%-------------------------------------------------------------
\subsection{Non-projectable potential terms}
\label{App:B4}
%-------------------------------------------------------------
Finally, we determine the heat-kernel coefficient of the 
interaction monomials including spatial derivatives of the lapse function.
Similarly to the case of the intrinsic curvatures, we limit ourselves 
to terms containing up to two derivatives of the component fields. These terms can be determined on a background $\cM = S^1 \times T^d$ with
the component fields restricted to the form
\be\label{back3}
N = N(x) \; , \quad N^i = 0 \; , \quad \sigma_{ij}(t,x) = \delta_{ij} \, ,
\ee
where the lapse $N(x)$ depends on the spatial coordinates $x$ only and the spatial metric is given by the Kronecker symbol. Setting $a_i = \p_i \ln N$, the two operators \eqref{DtDx} reduce to
\be
\Delta_t \equiv - N^{-2} \, \p_t^2 \, , \qquad \Delta_x = - \delta^{ij} \partial_i \partial_j - a^i \, \p_i \, , 
\ee
and the non-zero Christoffel symbols of the $D$-dimensional geometry are
\be\label{chris3}
\Gamma^i{}_{tt}= - N^2 a^i \; , \qquad \Gamma^t{}_{it} = a_i \, . 
\ee
Restricting the trace \eqref{optrace} to this background gives
\begin{equation}\label{eqb50}
\left. {\rm Tr} \, e^{-s D^2} \right|_{\cM} \simeq (4\pi)^{-(d+1)/2} \, s^{-\frac{1+d/z}{2}} \, \int  dt d^dx N  \left[ b_0 + c_1 \, s^{1/z} \, a_i a^i + \ldots \right] \, ,
\end{equation}
where the dots again indicate terms with more than two derivatives acting 
on the component fields.

The computation of the heat-kernel coefficient $c_1$ proceeds along the lines of App.\ \ref{sect.B3}. The explicit form of the commutators entering into \eqref{Tcommutators} is given by
\be\label{comtorus}
\begin{split}
\left[ \, \Delta_t \, , \, \Delta_x \, \right] = & - \left( 6 a_i a^i + 2 \left(\p_i a^i \right) + 4 a^i \partial_i \right) \, \Delta_t \, , \\
\left[ \, \Delta_t \, , \, \left[ \, \Delta_t \, , \, \Delta_x \, \right] \right] = & - 8 \, a_i a^i  \, \Delta_t^2 \, , \\ 
\left[ \, \Delta_x \, , \, \left[ \, \Delta_t \, , \, \Delta_x \, \right] \right] = & - 16 \, a^i a^j \, \p_i \p_j \, \Delta_t + \ldots \, , \\
\end{split}
\ee
where the dots again indicate terms which do not contribute to $c_1$. 
Since multi-commutators of higher
order automatically come with additional factors of $a^i$ or derivatives of $a^i$ the commutators \eqref{comtorus} are the only terms that contribute to the heat-kernel coefficient  $c_1$.

Abbreviating $I \equiv \int  dt d^dx N \, a_i a^i$  the explicit evaluation of the traces \eqref{Tcommutators} based on the commutators \eqref{comtorus} yields
\be
\begin{split}
T_1 \simeq & \, - \frac{11}{12}  (4\pi)^{-(d+1)/2} \, s^{-(d-2+z)/(2z)} \, \frac{1}{z} \,  \frac{\Gamma\left( \frac{d-2}{2z}\right)}{\Gamma\left(\frac{d-2}{2}\right)} \, I \, , \\ 	
T_2 \simeq & \, (4\pi)^{-(d+1)/2} \, s^{-(d-2+z)/(2z)} \, \left( 2 \, \frac{\Gamma\left( \frac{d-2}{2z} + 1 \right)}{\Gamma\left(\frac{d}{2}\right)}  + (z-1) \, \frac{\Gamma\left( \frac{d-2}{2z} + 1 \right)}{\Gamma\left(\frac{d+2}{2}\right)} \right) \, I \, , \\ 
T_3 \simeq & \, - \frac{3}{4} \, (4\pi)^{-(d+1)/2} \, s^{-(d-2+z)/(2z)} \, z \, \frac{\Gamma\left( \frac{d-2}{2z} + 2 \right)}{\Gamma\left(\frac{d+2}{2}\right)} \, I \, , \\
T_4 \simeq & \, - \frac{4}{3} \, (4\pi)^{-(d+1)/2} \, s^{-(d-2+z)/(2z)} \, z \, \frac{\Gamma\left( \frac{d-2}{2z} + 2 \right)}{\Gamma\left(\frac{d+2}{2}\right)} \, I \, , \\
T_5 \simeq & \, (4\pi)^{-(d+1)/2} \, s^{-(d-2+z)/(2z)} \,  \frac{\Gamma\left( \frac{d-2}{2z} + 1 \right)}{\Gamma\left(\frac{d}{2}\right)} \, I \, . 
\end{split}
\ee
Comparing the sum of these traces, 
\be
\left. {\rm Tr} \, e^{-sD^2} \right|_{I} = - \frac{13}{6}  (4\pi)^{-(d+1)/2} \, s^{-(d-2+z)/(2z)} \, \frac{z-1}{d} \, \frac{\Gamma\left(\frac{d-2}{2z}+1\right)}{\Gamma\left(\frac{d}{2}\right)} \, I \, , 
\ee
to the ansatz \eqref{eqb50} finally gives the heat-kernel coefficient
\be\label{e3coeff}
c_1 =  - \frac{13}{6} \, \frac{z-1}{d} \, \frac{\Gamma\left(\frac{d-2}{2z}+1\right)}{\Gamma\left(\frac{d}{2}\right)} \, . 
\ee
This result completes the derivation of the heat-kernel coefficients reported
in Sect.\ \ref{sect.3}.
%
%-------------------------------------------------------------
\end{appendix}
%-------------------------------------------------------------

%----- Bibliography ----------------------

%-----------------------------------------

%------------------------------------------------------------------------------
\end{document}